\documentclass[10pt,journal,compsoc]{IEEEtran}
\newcommand{\comment}[1]{}

\newif\ifcomment
\def\comment#1{%
    \ifcomment\relax\else #1\fi}
\commenttrue

\usepackage[dvipsnames]{xcolor}
\usepackage{multirow}

\usepackage{rotating}
\usepackage{colortbl}
\usepackage{tcolorbox}
\usepackage{tikz}
\usepackage{arydshln}
\usepackage{longtable}
\usepackage{scrextend}

\usepackage{makecell}

\usepackage{pifont}% http://ctan.org/pkg/pifont
\newcommand{\cmark}{\ding{51}}%
\newcommand{\xmark}{\ding{55}}%

\usepackage{caption}
\usepackage{array}
\newcolumntype{L}{>{\centering\arraybackslash}m{6cm}}
\newcolumntype{L}[1]{>{\raggedright\let\newline\\\arraybackslash\hspace{0pt}}m{#1}}
\newcolumntype{C}[1]{>{\centering\let\newline\\\arraybackslash\hspace{0pt}}m{#1}}
\newcolumntype{R}[1]{>{\raggedleft\let\newline\\\arraybackslash\hspace{0pt}}m{#1}}
\newcolumntype{P}[1]{>{\raggedright}p{#1}}

\usepackage{booktabs}

\renewcommand{\arraystretch}{1.8}

\definecolor{why}{HTML}{EE220C}
\definecolor{who}{HTML}{00A2FF}
\definecolor{what}{HTML}{1DB100}
\definecolor{how}{HTML}{F8BA00}
\definecolor{when}{HTML}{CB297B}
\definecolor{where}{HTML}{00A89D}

\definecolor{white}{HTML}{FFFFFF}
\definecolor{lightgray}{HTML}{F3F3F3}
\definecolor{black}{HTML}{000000}
\definecolor{tabletagcolor}{HTML}{BDBDBD}

\definecolor{cell}{HTML}{B0BEC5}
\definecolor{rowbackground}{HTML}{F0F2F4}
\newcommand{\f}{
\begin{tikzpicture}[every node/.style={inner sep=0,outer sep=0},scale=0.4]
    \fill [rounded corners=0.08cm,fill=cell] (0,0)--(.6,0)--(.6,.6)--(0,.6)--cycle;
\end{tikzpicture}
}

\newcommand{\rspace}{$\hspace{0.1cm}$}

\usepackage{tcolorbox}

\definecolor{tagbordercolor}{rgb}{0.8, 0.8, 0.8}
\definecolor{tagbgcolor}{rgb}{0.9, 0.9, 0.9}

\newtcbox{\whytag}{nobeforeafter, colframe=why,
colback=why, boxrule=0.5pt, arc=1pt,
 boxsep=0pt,left=2pt,right=2pt,top=1.5pt,bottom=2pt,tcbox raise base}

\newtcbox{\whotag}{nobeforeafter, colframe=who,
colback=who, boxrule=0.5pt, arc=1pt,
 boxsep=0pt,left=2pt,right=2pt,top=1.5pt,bottom=2pt,tcbox raise base}

\newtcbox{\whattag}{nobeforeafter, colframe=what,
colback=what, boxrule=0.5pt, arc=1pt,
 boxsep=0pt,left=2pt,right=2pt,top=1.5pt,bottom=2pt,tcbox raise base}

\newtcbox{\howtag}{nobeforeafter, colframe=how,
colback=how, boxrule=0.5pt, arc=1pt,
 boxsep=0pt,left=2pt,right=2pt,top=1.5pt,bottom=2pt,tcbox raise base}

\newtcbox{\whentag}{nobeforeafter, colframe=when,
colback=when, boxrule=0.5pt, arc=1pt,
 boxsep=0pt,left=2pt,right=2pt,top=1.5pt,bottom=2pt,tcbox raise base}

\newtcbox{\wheretag}{nobeforeafter, colframe=where,
colback=where, boxrule=0.5pt, arc=1pt,
 boxsep=0pt,left=2pt,right=2pt,top=1.5pt,bottom=2pt,tcbox raise base}

\newtcbox{\tabletag}{nobeforeafter, colframe=tabletagcolor,
colback=white, boxrule=0.5pt, arc=1pt,
 boxsep=0pt,left=2pt,right=2pt,top=1.5pt,bottom=2pt,tcbox raise base}

\DeclareMathAlphabet{\altmathcal}{OMS}{cmsy}{m}{n}
\usepackage{amsmath,bm}

\usepackage{enumitem}
\usepackage{xcolor}
\usepackage{amssymb}
\usepackage{scrextend}

\definecolor{metric-color}{HTML}{1eb300}
\definecolor{attack-color}{HTML}{ed230c}
\definecolor{defense-color}{HTML}{2c90fc}
\definecolor{application-color}{HTML}{9627f2}

\usepackage{graphicx}
\usepackage{caption}
\usepackage{subcaption}
\usepackage{float}
\ifCLASSOPTIONcompsoc
  \usepackage[nocompress]{cite}
\else
  \usepackage{cite}
\fi
\ifCLASSINFOpdf
\else
\fi
\begin{document}
\title{Graph Vulnerability and Robustness: A Survey}

\author{Scott~Freitas,
        Diyi~Yang,
        Srijan~Kumar,
        Hanghang~Tong,
        and~Duen~Horng~Chau% <-this % stops a space
\IEEEcompsocitemizethanks{
\IEEEcompsocthanksitem S. Freitas, D. Yang, S. Kumar and D.H. Chau is with the Department
of Computational Science and Engineering, Georgia Institute of Technology, Atlanta,
GA, 30313.
E-mail: \{safreita, diyi.yang, srijan, polo\}@gatech.edu

\IEEEcompsocthanksitem H. Tong is with the Department of Computer Science, University of Illinois at Urbana-Champaign, Urbana,
Illinois, 61801.
E-mail: htong@illinois.edu
}
% <-this % stops an unwanted space
% \thanks{Manuscript received April 15, 2021; revised April 15, 2021.}
}

\IEEEtitleabstractindextext{%
\begin{abstract}
The study of network robustness is a critical tool in the characterization and sense making of complex interconnected systems such as infrastructure, communication and social networks.
While significant research has been conducted in these areas, gaps in the surveying literature still exist. 
Answers to key questions are currently scattered across multiple scientific fields and numerous papers.
In this survey, we distill key findings across numerous domains and provide researchers crucial access to important information by---%
(1) summarizing and comparing recent and classical graph robustness measures; 
(2) exploring which robustness measures are most applicable to different categories of networks (e.g., social, infrastructure);
(3) reviewing common network attack strategies, and summarizing which attacks are most effective across different network topologies; and
(4) extensive discussion on selecting defense techniques to mitigate attacks across a variety of networks.
This survey guides researchers and practitioners in navigating the expansive field of network robustness, while summarizing answers to key questions.
We conclude by highlighting current research directions and open problems.
\end{abstract}

}

\maketitle

\IEEEdisplaynontitleabstractindextext

% \fi
\IEEEpeerreviewmaketitle

\IEEEraisesectionheading{\section{Introduction}\label{sec:introduction}}


\IEEEPARstart{T}{here} are three fundamental tasks in the study of network robustness: 
(i) development of measures to quantify network robustness, 
(ii) identification of network attack mechanisms, 
and (iii) construction of defensive techniques to resist network failures and recover from attacks. 
First mentioned as early as the 1970's~\cite{chvatal1973tough}, network robustness has a rich and storied history spanning numerous fields of engineering and science~\cite{klein1993resistance,beygelzimer2005improving, tong2010vulnerability,krishnamoorthy1987fault}.
This diversity of research has generated a variety of unique perspectives, providing fresh insight into challenging problems, while equipping researchers with fundamental knowledge for their investigations.
While the fields of study may be diverse, they are linked by a common definition of network robustness~\cite{ellens2013graph,chan2016optimizing,beygelzimer2005improving}:

\medskip

\begin{addmargin}[1em]{0em}
\textit{Robustness is a measure of a network's ability to continue functioning when part of the network is either naturally damaged or targeted for attack}.
\end{addmargin}

\medskip
\noindent
To provide some intuition for this definition, we consider an example of a power grid network that is susceptible to both \textit{natural} failures and \textit{targeted} attacks.
A natural failure happens when a \textit{single} power substation fails due to erosion of parts or natural disasters.
However, when one substation fails, additional load is routed to alternative substations, which can cause \textit{cascading failures}.
Not all failures originate from natural causes, some come from \textit{targeted} attacks, 
such as enemy states hacking into the grid to sabotage key equipment to maximally damage the operations of the electrical grid.
Through analyzing and understanding the robustness of these networks, we can mitigate damage from both natural failures and targeted attacks, and in some cases, prevent it altogether.

Unfortunately, the nature of cross-disciplinary research also comes with significant challenges.
Oftentimes important discoveries made in one field are not quickly disseminated, leading to missed innovation opportunities.
In this survey, our goal is to distill key research questions raised in prior related research~\cite{chan2016optimizing}, that if addressed effectively, will assist readers in understanding the complex interconnections of this topic, and accelerate the dissemination of ideas. 
Specifically, we analyze and compare numerous classical and modern robustness techniques---addressing a crucial gap in the survey literature, and helping set the stage for future work to be built upon.

\subsection{Contributions}
We distill and summarize critical topics found in the literature of network robustness through contributions C1-C5.

\vspace{0.2cm}
\noindent \textbf{C1. Summary of Robustness Measures}
We summarize 17 modern and classic network robustness measures, along with how each measure is \textit{linked} to the evaluation of graph vulnerability and robustness.
Our goal is to provide researchers a repository of information to objectively compare robustness measures for use in their own applications.

\begin{figure*}[t!]
    \centering
    \includegraphics[width=\textwidth]{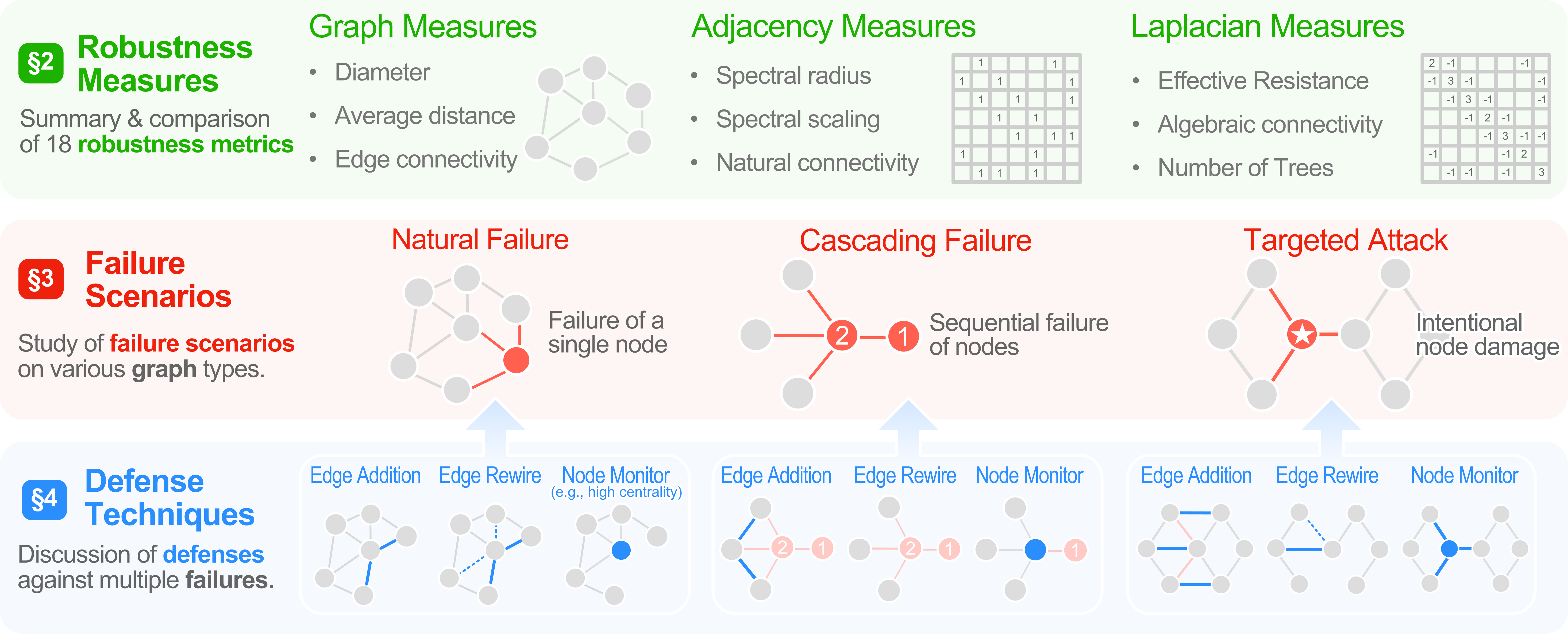}
    \caption{A visual overview of the work surveyed in this paper. §2 summarizes and compares 17 graph robustness measures. 
    §3 overviews methods of network failure and attack.
    §4 summarizes network defense techniques across a variety of graph topologies and attack vectors.
    %\textcolor{metric-color}{§2}
    %\textcolor{attack-color}{§3}
    %\textcolor{defense-color}{§4}
    }
    \label{fig:crown}
\end{figure*}

\vspace{0.2cm}
\noindent\textbf{C2. Exploration of Robustness Measure Applications}
Not all robustness measures are equally applicable to every category of network data.
For example, the graph clustering coefficient is important to the study of social networks since it provides an indication of group ``tightness''.
However, water distribution networks require measures that can account for bottlenecks and alternative pathways.
We delve into the literature and summarize why some measures are more applicable to particular domains. 
In particular, we study two high-impact use case scenarios---(i) transportation networks and (ii) water distribution networks.

\vspace{0.2cm}
\noindent \textbf{C3. Overview of Network Attack Strategies}
By understanding the topology of the network, we can analyze the effects of natural failures and targeted attacks.
We discuss popular network attack strategies, and provide a high level summary of which attack strategies are best adapted for different network topologies.
This attack strategy comparison, as the first of its kind, provides researchers and practitioners a quick guide to effective attack selection. 

\vspace{0.2cm}
\noindent \textbf{C4. Comparison of Network Defense Mechanisms}
By understanding the common network topologies and attack strategies from C3, we can study the effectiveness of network defense mechanisms.
We summarize numerous defense mechanisms in relation to various network topologies, mechanisms of natural failure, and targeted attacks across the three primary mechanisms for defending a network:

\begin{enumerate}[label=(\roman*), labelindent=3mm, align=left, topsep=6pt, leftmargin=*, itemsep=4pt, widest=iii]
    \item \textit{edge rewiring} (e.g., rewire power lines) 
    
    \item \textit{edge addition} (e.g., add additional power lines)
    
    \item \textit{node monitoring} (e.g., closely monitor substation)
\end{enumerate}
 
\noindent Each mechanism has an associated benefit and cost, where some are disproportionately more expensive than others.
We explore these trade-offs with the goal of providing the reader a comprehensive overview of available defense options across a variety of real world scenarios to assist in the decision making of network defense.
To the best of our knowledge, this is the first comprehensive comparison of network defense techniques across attack vectors and network topology.

\vspace{0.2cm}
\noindent \textbf{C5. Highlight Open Problems and Research Directions}
Through careful analysis of the existing network robustness literature, we identify and distill open problems that have strong potential as future research directions.
Promising directions and open problems for future network robustness research include---%
(1) an axiomatic study of desired properties in robustness measures, helping guide the selection and development of new measures;
(2) interpretability of robustness measures to assist users in understanding the impact of measure scores;
(3) applying the study of network robustness to additional high-impact domains such as physical security and cybersecurity;
and (4) bridging the study of graph vulnerability and robustness with recent developments in adversarial machine learning on graph structured data.

\subsection{Survey Methodology \& Summarization Process}
We study an extensive number of existing works and identify three main types of research contributions that they aim to make: 
(i) study of network robustness measures;
(ii) evaluation and development of attacks; 
and 
(iii) evaluation and development of defense mechanisms.
We frame our work based on these contributions.
Doing so allows us to 
summarize and compare relevant works from computer science, engineering, mathematics, and the sciences. 
This helps equip researchers with fundamental knowledge for their investigation, and provide them with fresh insights.
Specifically, we select works from top journals and conferences from the relevant domains.
Table~\ref{table:venues} lists some of the most prominent publication venues and their acronyms.
We also include papers posted on arXiv, an open-access, electronic repository, as many cutting-edge papers are first released here.

As the study of graph robustness has been carried out in a variety of fields (e.g., mathematics, physics, computer science), the terminology often varies from field to field. 
As such, we refer to the following word pairs interchangeably: (network, graph), (vertex, node), (edge, link), (adversary, attacker).
We follow standard notation and use capital bold letters for matrices (e.g., $\bm{A}$), lower-case bold letters for vectors (e.g., $\bm{a}$) and calligraphic font for sets (e.g., $\altmathcal{S}$). 
Throughout this paper we focus our attention on undirected and unweighted graphs, unless otherwise noted.
The reader may want to refer to Table~\ref{table:notation} throughout the survey for technical terms.

\begin{table}[H]
\footnotesize
\centering
\sffamily
\setlength{\tabcolsep}{1.25pt}
\renewcommand\arraystretch{1.3}

\begin{tabular}{p{1.45cm} p{6.6cm}} 
\toprule
    Nature & Nature \\ % 42.778
    PR & Physics Reports: A Review Section of Physics Letters \\ % 28.295
    Smart Grid & IEEE Transactions on Smart Grid \\ % 10.49
    PNAS & National Academy of Sciences of the USA \\ % 9.412
    PRL & Physical Review Letters \\ % 8.385
    SIREV & SIAM Review \\ % 8.02
    SMC & IEEE Systems, Man, and Cybernetics: Systems \\ % 5.131
    TKDE & IEEE Transactions on Knowledge \& Data Engineering \\ % 4.935
    CNSNS & Communications in Nonlinear Science and Numerical Simulation \\ % 4.115
    KAIS & Knowledge and Information Systems \\ % 2.936
    Physica A & Physica A: Statistical Mechanics and its Applications \\ % 2.924
    RA & Risk Analysis \\ % 2.898
    CHAOS & An Interdisciplinary Journal of Nonlinear Science \\ % 2.832
    PLOS & PLOS ONE \\ % 2.740
    DMKD & Data Mining and Knowledge Discovery \\ % 2.629
    SN & Social Networks: An International Journal of Structural Analysis \\[0.25em] % 2.376
    PRE & Physical Review E \\ % 2.296
    TOPS & ACM Transactions on Privacy and Security \\ % 1.974
    EPL & A Letters Journal Exploring the Frontiers of Physics \\ % 1.957
    Physica B & Physica B: Condensed Matter \\ % 1.902
    JOCN & Journal of Complex Networks \\ % 1.83
    EPJ B & The European Physical Journal B \\ % 1.440
    CPL & Chinese Physics Letters \\ % 1.066
    FCS & Frontiers of Computer Science \\ % 1.039
    LAA & Linear Algebra and its Applications \\ % 0.972

    \cdashline{1-2}[.7pt/1.5pt]\noalign{\vskip 0.25em}
    Web & The Web Conference (formerly WWW) \\ % 14.69
    KDD & ACM Knowledge Discovery \& Data Mining \\ % 12.04
    WSDM & ACM Conference on Web Search \& Data Mining \\ % 10.56
    ICDM & IEEE International Conference on Data Mining \\ % 2.32
    CIKM & ACM Information \& Knowledge Management \\ % 9.4
    NDSS & Network and Distributed System Security Symposium \\ % 8.91
    INFO & IEEE Conference on Computer Communications \\ % 7.29
    CDC & IEEE Conference on Decision and Control \\ % 6.11
    SDM & SIAM International Conference on Data Mining \\ % 5.45
    PKDD &  European Conference on Machine Learning and Principles and Practice of Knowledge Discovery in Databases \\[0.25em] % 4.95
    PAKDD & Pacific-Asia Conference on Knowledge Discovery \& Data Mining \\ % 3.857
    A-POL & Transportation Research Part A: Policy and Practice \\ % 3.693
    GLOBE & IEEE Global Communications Conference \\ % 2.15
    DRCN & Design of Reliable Communication Networks \\ % 1.83

    \cdashline{1-2}[.7pt/1.5pt]\noalign{\vskip 0.25em}

    RNDM & Resilient Networks Design and Modeling Workshop \\ % 2.05
    WORM & ACM Workshop on Rapid Malcode \\
    SIMPLEX &  Simplifying Complex Network for Practitioners \\[0.25em]
    
    \cdashline{1-2}[.7pt/1.5pt]\noalign{\vskip 0.25em}
    
    arXiv & arXiv.org e-Print Archive \\
    
\bottomrule
\end{tabular}
\caption{Relevant venues in order of journals, conferences, workshops, and preprints.
Within each category, we order the venues based on the most recently available impact factors reported officially (e.g., venues' websites).
}
\label{table:venues}
\end{table}

\subsection{Related Surveys}
While many surveys have been conducted in domain specific applications of network robustness, including water distribution networks~\cite{yazdani2012applying}, airline route networks~\cite{lordan2014study}, power grids~\cite{pagani2013power,cuadra2015critical}, to our knowledge there is no survey that summarizes the graph robustness landscape at large.
Different from all the related articles mentioned above, our survey provides a comprehensive, cross-domain framework to describe graph vulnerability and robustness, discusses the rapidly growing community at large, and presents major research trajectories synthesized from existing literature.

\begin{table*}[ht!]
\footnotesize
\sffamily
\centering

\setlength{\tabcolsep}{1.25pt}
\renewcommand\arraystretch{1.2}
\begin{tabular}{
% paper
R{2.9cm}|
% metrics
C{0.51cm}C{0.51cm}C{0.51cm}C{0.51cm}C{0.51cm}C{0.51cm}C{0.51cm}C{0.51cm}C{0.51cm}:C{0.51cm}C{0.51cm}C{0.51cm}C{0.51cm}C{0.51cm}:C{0.51cm}C{0.51cm}C{0.51cm}|
% attacks
C{0.51cm}C{0.51cm}|
% defense
C{0.51cm}C{0.51cm}C{0.51cm}|
% application
% C{0.51cm}C{0.51cm}C{0.51cm}C{0.51cm}|
% where
L{1.6cm}
}
% titles
% blank above paper cell
\multicolumn{1}{c}{} &

\multicolumn{17}{c}{{\normalsize{\textbf{\textcolor{metric-color}{Robustness Measures}}}}} &

\multicolumn{2}{c}{\normalsize{\textbf{{\textcolor{attack-color}{Attack}}}}} &

\multicolumn{3}{c}{\normalsize{\textbf{{\textcolor{defense-color}{Defense}}}}} &

% \multicolumn{4}{c}{\normalsize{\textbf{{\textcolor{application-color}{Network}}}}} &

\multicolumn{1}{c}{\normalsize{\textbf{{\textcolor{gray}{Where}}}}} \\

\cmidrule(l){2-18}\cmidrule(l){19-20}\cmidrule(l){21-23}\cmidrule(l){24-24}

% \multicolumn{1}{c}{} &
% \multicolumn{7}{c}{{\normalsize{\textbf{$G$}}}} &
% \multicolumn{3}{c}{{\normalsize{\textbf{$A$}}}} &
% \multicolumn{3}{c}{{\normalsize{\textbf{$L$}}}} \\

% paper
\multicolumn{1}{r}{\textbf{Work} \rspace} & 

% metrics
\multicolumn{1}{c}{\rotatebox{90}{\tabletag{ \textbf{\ref{subsec:binary-connectivity}} } Binary Connnectivity}} &
\multicolumn{1}{c}{\rotatebox{90}{\tabletag{ \textbf{\ref{subsec:vertex-connectivity}} } Vertex Connectivity}}  &
\multicolumn{1}{c}{\rotatebox{90}{\tabletag{ \textbf{\ref{subsec:edge-connectivity}} } Edge Connectivity}}  &
\multicolumn{1}{c}{\rotatebox{90}{\tabletag{ \textbf{\ref{subsec:diameter}} } Diameter}}  &
\multicolumn{1}{c}{\rotatebox{90}{\tabletag{ \textbf{\ref{subsec:average-distance}} } Average Distance}}  &
\multicolumn{1}{c}{\rotatebox{90}{\tabletag{ \textbf{\ref{subsec:vertex-betweenness}} } Avg. Vertex Betweenness}}  &
\multicolumn{1}{c}{\rotatebox{90}{\tabletag{ \textbf{\ref{subsec:edge-betweenness}} } Avg. Edge Betweenness}}  &
\multicolumn{1}{c}{\rotatebox{90}{\tabletag{ \textbf{\ref{subsec:gcc}} } Global Clustering Coefficient}}  &
\multicolumn{1}{c:}{\rotatebox{90}{\tabletag{ \textbf{\ref{subsec:lcc}} } Largest Connected Component}}  &

\multicolumn{1}{c}{\rotatebox{90}{\tabletag{ \textbf{\ref{subsec:spectral-radius}} } Spectral Radius}}  &
\multicolumn{1}{c}{\rotatebox{90}{\tabletag{ \textbf{\ref{subsec:spectral-gap}} } Spectral Gap}}  &
\multicolumn{1}{c}{\rotatebox{90}{\tabletag{ \textbf{\ref{subsec:natural-connectivity}} } Natural Connectivity}}  &
\multicolumn{1}{c}{\rotatebox{90}{\tabletag{ \textbf{\ref{subsec:spectral-scaling}} } Spectral Scaling}}  &
\multicolumn{1}{c:}{\rotatebox{90}{\tabletag{ \textbf{\ref{subsec:robustness-index}} } Generalized Robustness Index}}  &

\multicolumn{1}{c}{\rotatebox{90}{\tabletag{ \textbf{\ref{subsec:algebraic-connectivity}} } Algebraic Connectivity}}  &
\multicolumn{1}{c}{\rotatebox{90}{\tabletag{ \textbf{\ref{subsec:spanning-trees}} } Number of Spanning Trees}}  &
\multicolumn{1}{c|}{\rotatebox{90}{\tabletag{ \textbf{\ref{subsec:effective-resistance}} } Effective Resistance}}  &

% attacks
\multicolumn{1}{c}{\rotatebox{90}{\tabletag{ \textbf{\ref{subsec:targeted-attacks}} } Node Removal}} &
\multicolumn{1}{c|}{\rotatebox{90}{\tabletag{ \textbf{\ref{subsec:targeted-attacks}} } Edge Removal}} &

% Defense
\multicolumn{1}{c}{\rotatebox{90}{\tabletag{ \textbf{\ref{subsubsec:edge-addition}} } Edge Addition}} &
\multicolumn{1}{c}{\rotatebox{90}{\tabletag{ \textbf{\ref{subsubsec:edge-rewiring}} } Edge Rewiring}} &
\multicolumn{1}{c|}{\rotatebox{90}{\tabletag{ \textbf{\ref{subsubsec:node-monitoring}} } Node Monitoring}} &

% where
\multicolumn{1}{c}{\rspace \rotatebox{90}{Publication Venue}} \\

\midrule

% papers
\rowcolor{rowbackground}
Albert,et.al.\cite{albert2000error} \rspace  &  &  &  & \f & \f &  &  &  & \f &  &  &  &  &  &  &  &  & \f &  &  &  &  & \rspace Nature \\

Alenazi,et.al.\cite{alenazi2015evaluation} \rspace  &  &  &  &  & \f & \f &  & \f &  &  & \f & \f &  &  & \f &  &  & \f &  & \f &  &  & \rspace RNDM \\

\rowcolor{rowbackground}
Alenazi,et.al.\cite{alenazi2015comprehensive} \rspace  &  &  &  &  & \f & \f &  & \f &  &  &  &  & \f & \f & \f &  & \f &  &  &  &  &  & \rspace DRCN \\

Baig,et.al.\cite{baig2015correlation} \rspace  &  &  &  &  &  &  &  &  &  &  &  &  &  &  &  &  &  & \f &  &  &  &  &  \rspace Web \\

\rowcolor{rowbackground}
Baras,et.al.\cite{baras2009efficient} \rspace  &  &  &  &  &  &  &  &  &  &  &  &  &  &  & \f & \f & \f &  &  & \f &  &  &  \rspace CDC \\

Berdica\cite{berdica2002introduction} \rspace  &  &  & \f &  &  &  &  &  &  &  &  &  &  &  &  &  &  &  &  &  &  &  &  \rspace A-POL \\

\rowcolor{rowbackground}
Bernstein,et.al.\cite{bernstein2014power} \rspace  &  &  &  &  &  &  &  &  &  &  &  &  &  &  &  &  &  &  & \f &  &  &  &  \rspace INFO \\

Beygelzimer,et.al.\cite{beygelzimer2005improving} \rspace  &  &  &  & \f & \f &  &  &  & \f &  &  &  &  &  &  &  &  & \f &  & \f & \f &  &  \rspace Physica A \\

\rowcolor{rowbackground}
Bigdeli,et.al.\cite{bigdeli2009comparison} \rspace  &  &  &  &  &  & \f &  &  &  &  &  &  &  &  & \f &  &  &  &  &  &  &  &  \rspace SIMPLEX \\

Bishop,et.al.\cite{bishop2011link} \rspace  &  &  &  &  &  &  &  &  &  & \f &  &  &  &  &  &  &  &  & \f &  &  &  & \rspace EPL \\

\rowcolor{rowbackground}
Bocca,et.al.\cite{boccaletti2006complex} \rspace  &  &  &  & \f & \f & \f & \f & \f & \f &  &  &  &  &  & \f &  &  & \f & \f &  &  &  &  \rspace PR \\

Borgatti,et.al.\cite{borgatti2006robustness} \rspace  &  &  &  &  &  & \f &  &  &  &  &  &  &  &  &  &  &  & \f & \f & \f &  &  & \rspace SN \\

\rowcolor{rowbackground}
Briesemeis.,et.al.\cite{briesemeister2003epidemic} \rspace  &  &  &  &  &  &  &  &  &  &  &  &  &  &  &  &  &  &  &  &  &  & \f & \rspace WORM \\

Buldyrev,et.al.\cite{buldyrev2010catastrophic} \rspace  &  &  &  &  &  &  &  &  & \f &  &  &  &  &  &  &  &  & \f &  &  &  &  &  \rspace Nature \\

\rowcolor{rowbackground}
Byrne,et.al.\cite{byrne2005algebraic} \rspace  &  & \f & \f &  & \f &  &  &  &  &  &  &  &  &  & \f &  &  &  &  &  &  &  & \rspace Sandia \\

Caballero,et.al.\cite{caballero2008would} \rspace  &  &  &  &  &  &  &  &  & \f &  &  &  &  &  &  &  &  &  & \f &  &  &  & \rspace NDSS \\

\rowcolor{rowbackground}
Callaway,et.al.\cite{callaway2000network} \rspace  &  &  &  &  &  &  &  &  & \f &  &  &  &  &  &  &  &  & \f &  &  &  &  & \rspace PRL \\

Chan,et.al.\cite{chan2014make} \rspace  &  &  &  &  &  &  &  &  &  &  &  & \f &  &  &  &  &  & \f & \f &  &  &  & \rspace SDM \\

\rowcolor{rowbackground}
Chakrabarti,et.al.\cite{chakrabarti2008epidemic} \rspace  &  &  &  &  &  &  &  &  &  & \f &  &  &  &  &  &  &  &  &  &  &  & \f &  \rspace TOPS \\

Chan,et.al.\cite{chan2016optimizing} \rspace  &  &  &  &  &  &  &  &  &  &  & \f & \f & \f &  &  \f & \f & \f &  &  &  & \f &  & \rspace DMKD \\

\rowcolor{rowbackground}
Chen,et.al.\cite{chen2015node} \rspace  &  &  &  &  &  &  &  &  &  & \f &  &  &  &  &  &  &  &  &  &  &  & \f & \rspace TKDE \\

Chen,et.al.\cite{chen2015connectivity} \rspace  &  &  &  &  &  &  &  & \f &  & \f &  &  &  &  &  &  &  & \f &  &  &  &  &  \rspace ICDM \\

\rowcolor{rowbackground}
Chen,et.al.\cite{chen2016eigen} \rspace  &  &  &  &  &  &  &  &  &  & \f &  &  &  &  &  &  &  &  & \f & \f &  &  &  \rspace TKDD \\

Crucitti,et.al.\cite{crucitti2004model} \rspace  &  &  &  &  & \f &  &  &  &  &  &  &  &  &  &  &  &  & \f &  &  &  &  &  \rspace PRE \\

\rowcolor{rowbackground}
Dekker\cite{dekker2005simulating} \rspace  &  & \f &  & \f &  &  &  &  &  &  &  &  &  &  &  &  &  & \f &  &  &  &  &  \rspace ACSC \\

Derrible,et.al.\cite{derrible2010complexity} \rspace  &  &  &  & \f & \f &  &  & \f &  &  &  &  &  &  &  &  &  &  &  &  &  &  &  \rspace Physica A \\

\rowcolor{rowbackground}
Duan,et.al.\cite{duan2014robustness} \rspace  &  &  &  &  & \f & \f &  & \f & \f &  &  &  &  &  &  &  &  & \f &  &  &  &  &  \rspace Physica A \\

Ellens,et.al.\cite{ellens2011effective} \rspace  &  &  &  &  &  &  &  &  &  &  &  &  &  &  &  &  & \f &  &  & \f &  &  &  \rspace LAA \\

\rowcolor{rowbackground}
Ellens,et.al.\cite{ellens2013graph} \rspace & \f & \f & \f & \f & \f & \f & \f & \f &  &  &  &  &  &   & \f & \f & \f &  &  &  &  &  &  \rspace arXiv \\

Estrada,et.al.\cite{estrada2006network} \rspace  &  &  &  &  &  &  &  &  &  &  & \f &  & \f &  &  &  &  &  &  &  &  &  &  \rspace Physica B \\

\rowcolor{rowbackground}
Estrada,et.al.\cite{estrada2006spectral} \rspace  &  &  &  &  &  &  &  &  &  &  & \f &  & \f &  &  &  &  &  &  &  &  &  &  \rspace EPL \\

Freitas,et.al.\cite{freitas2020d2m} \rspace  &  &  &  &  &  &  &  &  &  &  &  &  &  &  &  &  &  &  &  &  &  & \f & \rspace SDM \\

\rowcolor{rowbackground}
Freitas,et.al.\cite{freitas2021evaluating} \rspace  & \f & \f & \f & \f & \f & \f & \f & \f & \f & \f & \f & \f & \f & \f & \f & \f & \f & \f & \f & \f & \f & \f & \rspace CIKM \\

Gao,et.al.\cite{gao2011robustness} \rspace  &  &  &  &  &  &  &  &  & \f &  &  &  &  &  &  &  &  & \f &  &  &  &  &  \rspace PRL \\

\rowcolor{rowbackground}
Ghosh,et.al.\cite{ghosh2008minimizing} \rspace  &  &  &  &  &  &  &  &  &  &  &  &  &  &  &  &  & \f &  &  &  &  &  & \rspace SIREV \\

Holme,et.al.\cite{holme2002attack} \rspace  &  &  &  &  & \f &  &  &  & \f &  &  &  &  &  &  &  &  & \f & \f &  &  &  & \rspace PRE \\

\rowcolor{rowbackground}
Holmgren\cite{holmgren2006using} \rspace  &  &  &  &  & \f &  &  & \f & \f &  &  &  &  &  &  &  &  & \f &  &  &  &  &  \rspace RA \\

Jamakovic,et.al.\cite{jamakovic2007relationship} \rspace  &  & \f & \f &  &  &  &  &  &  &  &  &  &  &  & \f &  &  & \f & \f &  &  &  &  \rspace NGI \\

\rowcolor{rowbackground}
khalil,et.al.\cite{khalil2014scalable} \rspace  &  &  &  &  &  &  & \f &  &  & \f &  &  &  &  &  &  &  &  & \f & \f &  &  & \rspace KDD \\

Kinney,et.al.\cite{kinney2005modeling} \rspace  &  &  &  &  & \f &  &  &  &  &  &  &  &  &  &  &  &  & \f &  &  &  &  &  \rspace EPJ B \\

\rowcolor{rowbackground}
Klau,et.al.\cite{klau2005robustness} \rspace  & \f & \f &  & \f & \f &  &  &  & \f &  &  &  &  &  &  &  &  & \f & \f &  &  &  & \rspace Net. Anal. \\

Latora,et.al.\cite{latora2005vulnerability} \rspace  &  &  &  &  & \f &  &  &  &  &  &  &  &  &  &  &  &  & \f & \f & \f &  &  &  \rspace PRE \\

\rowcolor{rowbackground}
Le,et.al.\cite{le2015met} \rspace  &  &  &  &  &  &  &  &  &  &  &  &  &  &  &  &  &  &  & \f &  &  &  &  \rspace SDM \\

Leskovec,et.al.\cite{leskovec2007cost} \rspace  &  &  &  &  &  &  &  &  &  &  &  &  &  &  &  &  &  &  &  &  &  & \f &  \rspace KDD \\

\rowcolor{rowbackground}
Liu,et.al.\cite{liu2017comparative} \rspace  &  &  &  &  & \f & \f & \f &  &  &  &  & \f &  &  & \f &  &  & \f & \f &  &  &  & \rspace FCS \\

Lu,et.al.\cite{lu2016attack} \rspace  &  &  &  &  &  & \f & \f &  &  &  &  &  &  &  &  &  &  & \f & \f &  &  &  &  \rspace PLOS One \\

\rowcolor{rowbackground}
Malliaros,et.al.\cite{malliaros2012fast} \rspace  &  &  &  &  &  &  &  &  &  &  &  &  &  &  \f &  &  &  &  &  &  &  &  &  \rspace SDM \\

\midrule

\end{tabular}
\label{table:papers}
\end{table*}

\begin{table*}[ht!]
\footnotesize
\sffamily
\centering

\setlength{\tabcolsep}{1.25pt}
\renewcommand\arraystretch{1.2}
\begin{tabular}{
% paper
R{2.9cm}|
% metrics
C{0.51cm}C{0.51cm}C{0.51cm}C{0.51cm}C{0.51cm}C{0.51cm}C{0.51cm}C{0.51cm}C{0.51cm}:C{0.51cm}C{0.51cm}C{0.51cm}C{0.51cm}C{0.51cm}:C{0.51cm}C{0.51cm}C{0.51cm}|
% attacks
C{0.51cm}C{0.51cm}|
% defense
C{0.51cm}C{0.51cm}C{0.51cm}|
% application
% C{0.51cm}C{0.51cm}C{0.51cm}C{0.51cm}|
% where
L{1.6cm}
}
% titles
% blank above paper cell
% \multicolumn{1}{c}{} &

% \multicolumn{17}{c}{{\normalsize{\textbf{\textcolor{metric-color}{Robustness Measures}}}}} &

% \multicolumn{2}{c}{\normalsize{\textbf{{\textcolor{attack-color}{Attack}}}}} &

% \multicolumn{3}{c}{\normalsize{\textbf{{\textcolor{defense-color}{Defense}}}}} &

% % \multicolumn{4}{c}{\normalsize{\textbf{{\textcolor{application-color}{Network}}}}} &

% \multicolumn{1}{c}{\normalsize{\textbf{{\textcolor{gray}{Where}}}}} \\

% \cmidrule(l){2-18}\cmidrule(l){19-20}\cmidrule(l){21-23}\cmidrule(l){24-24}

% % \multicolumn{1}{c}{} &
% % \multicolumn{7}{c}{{\normalsize{\textbf{$G$}}}} &
% % \multicolumn{3}{c}{{\normalsize{\textbf{$A$}}}} &
% % \multicolumn{3}{c}{{\normalsize{\textbf{$L$}}}} \\

% paper
\multicolumn{1}{r}{\textbf{Work} \rspace} & 

% metrics
\multicolumn{1}{c}{\rotatebox{90}{\tabletag{ \textbf{\ref{subsec:binary-connectivity}}}}} &
\multicolumn{1}{c}{\rotatebox{90}{\tabletag{ \textbf{\ref{subsec:vertex-connectivity}} }}}  &
\multicolumn{1}{c}{\rotatebox{90}{\tabletag{ \textbf{\ref{subsec:edge-connectivity}} }}}  &
\multicolumn{1}{c}{\rotatebox{90}{\tabletag{ \textbf{\ref{subsec:diameter}}}}}  &
\multicolumn{1}{c}{\rotatebox{90}{\tabletag{ \textbf{\ref{subsec:average-distance}}}}}  &
\multicolumn{1}{c}{\rotatebox{90}{\tabletag{ \textbf{\ref{subsec:vertex-betweenness}}}}}  &
\multicolumn{1}{c}{\rotatebox{90}{\tabletag{ \textbf{\ref{subsec:edge-betweenness}}}}}  &
\multicolumn{1}{c}{\rotatebox{90}{\tabletag{ \textbf{\ref{subsec:gcc}}}}}  &
\multicolumn{1}{c}{\rotatebox{90}{\tabletag{ \textbf{\ref{subsec:lcc}}}}}  &

\multicolumn{1}{c}{\rotatebox{90}{\tabletag{ \textbf{\ref{subsec:spectral-radius}}}}}  &
\multicolumn{1}{c}{\rotatebox{90}{\tabletag{ \textbf{\ref{subsec:spectral-gap}}}}}  &
\multicolumn{1}{c}{\rotatebox{90}{\tabletag{ \textbf{\ref{subsec:natural-connectivity}}}}}  &
\multicolumn{1}{c}{\rotatebox{90}{\tabletag{ \textbf{\ref{subsec:spectral-scaling}}}}}  &
\multicolumn{1}{c}{\rotatebox{90}{\tabletag{ \textbf{\ref{subsec:robustness-index}}}}}  &

\multicolumn{1}{c}{\rotatebox{90}{\tabletag{ \textbf{\ref{subsec:algebraic-connectivity}}}}}  &
\multicolumn{1}{c}{\rotatebox{90}{\tabletag{ \textbf{\ref{subsec:spanning-trees}}}}}  &
\multicolumn{1}{c}{\rotatebox{90}{\tabletag{ \textbf{\ref{subsec:effective-resistance}}}}}  &

% attacks
\multicolumn{1}{c}{\rotatebox{90}{\tabletag{ \textbf{\ref{subsec:targeted-attacks}}}}} &
\multicolumn{1}{c}{\rotatebox{90}{\tabletag{ \textbf{\ref{subsec:targeted-attacks}}}}} &

% Defense
\multicolumn{1}{c}{\rotatebox{90}{\tabletag{ \textbf{\ref{subsubsec:edge-addition}}}}} &
\multicolumn{1}{c}{\rotatebox{90}{\tabletag{ \textbf{\ref{subsubsec:edge-rewiring}}}}} &
\multicolumn{1}{c}{\rotatebox{90}{\tabletag{ \textbf{\ref{subsubsec:node-monitoring}}}}} &

% where
\multicolumn{1}{c}{Venue} \\

\midrule

% papers

Marzo,et.al.\cite{marzo2018selecting} \rspace  &  & \f & \f & \f & \f & \f & \f & \f & \f & \f &  & \f &  &  & \f & \f & \f &  &  &  &  &  &  \rspace RNDM \\

\rowcolor{rowbackground}
Mattsson,et.al.\cite{mattsson2015vulnerability} \rspace  &  &  &  &  & \f & \f & \f &  & \f &  &  &  &  &  &  &  &  & \f & \f &  &  &  &  \rspace A-POL \\

Mieghem,et.al.\cite{mieghem2011decreasing} \rspace  &  &  &  &  &  &  &  &  & \f &  &  &  &  &  &  &  &  & \f & \f &  &  &  &  \rspace PRE \\

\rowcolor{rowbackground}
Milanese,et.al.\cite{milanese2010approximating} \rspace  &  &  &  &  &  &  &  &  &  & \f & \f &  &  &  &  &  &  &  & \f & \f &  &  & \rspace PRE \\

Motter,et.al.\cite{motter2002cascade} \rspace  &  &  &  &  &  &  &  &  & \f &  &  &  &  &  &  &  &  & \f &  &  &  &  &  \rspace PRE \\

\rowcolor{rowbackground}
Nardo,et.al.\cite{nardo2018applications} \rspace  &  &  &  &  &  &  &  &  &  & \f & \f &  &  &  & \f &  &  &  &  &  &  &  &  \rspace Water \\

Nguyen,et.al.\cite{nguyen2013detecting} \rspace  &  &  &  &  &  & \f &  &  & \f &  &  &  &  &  &  &  &  & \f &  &  &  &  &  \rspace Smart Grid \\

\rowcolor{rowbackground}
Parandeh.,et.al.\cite{parandehgheibi2013robustness} \rspace  &  &  &  &  &  &  &  &  &  &  &  &  &  &  &  &  &  & \f &  &  &  &  & \rspace GLOBE \\

Parshani,et.al.\cite{parshani2010interdependent} \rspace  &  &  &  &  &  &  &  &  & \f &  &  &  &  &  &  &  &  & \f &  &  &  &  & \rspace  PRL \\

\rowcolor{rowbackground}
Paul,et.al.\cite{paul2004optimization} \rspace  &  &  &  &  &  &  &  &  &  &  &  &  &  &  &  &  &  & \f &  &  &  &  & \rspace EPJ B \\

Prakash,et.al.\cite{prakash2010virus} \rspace  &  &  &  &  &  &  &  &  &  & \f &  &  &  &  &  &  &  &  &  &  &  & \f &  \rspace PKDD \\

\rowcolor{rowbackground}
Prakash,et.al.\cite{prakash2012threshold} \rspace  &  &  &  &  &  &  &  &  &  & \f &  &  &  &  &  &  &  &  &  &  &  &  & \rspace KAIS \\

Prakash,et.al.\cite{prakash2013fractional} \rspace  &  &  &  &  &  &  &  &  &  &  & \f &  &  &  &  &  &  &  &  &  &  & \f &  \rspace SDM \\

\rowcolor{rowbackground}
Rueda,et.al.\cite{rueda2017robustness} \rspace  &  & \f & \f & \f & \f & \f & \f & \f &  & \f &  & \f &  &  &  & \f & \f & \f & \f &  &  &  & \rspace JNSM \\

Saha,et.al.\cite{saha2015approximation} \rspace  &  &  &  &  &  &  &  &  &  & \f &  &  &  &  &  &  &  & \f & \f &  &  &  &  \rspace SDM \\

\rowcolor{rowbackground}
Schneider,et.al.\cite{schneider2011mitigation} \rspace  &  &  &  &  &  &  &  &  & \f &  &  &  &  &  &  &  &  &  &  &  & \f &  & \rspace PNAS \\

Schneider,et.al.\cite{schneider2011suppressing} \rspace  &  &  &  &  &  & \f & \f &  & \f &  &  &  &  &  &  &  &  &  &  &  &  & \f & \rspace PRE \\

\rowcolor{rowbackground}
Shao,et.al.\cite{shao2011cascade} \rspace  &  &  &  &  &  &  &  &  & \f &  &  &  &  &  &  &  &  & \f &  &  &  &  & \rspace PRE \\

Shargel,et.al.\cite{shargel2003optimization} \rspace  &  &  &  &  &  &  &  &  &  &  &  &  &  &  &  &  &  & \f &  &  &  &  & \rspace PRL \\

\rowcolor{rowbackground}
Sydney,et.al.\cite{sydney2008elasticity} \rspace  &  &  &  &  &  &  &  &  &  &  &  &  &  &  & \f &  &  & \f & \f &  &  &  & \rspace arXiv \\

Tanaka,et.al.\cite{tanaka2012dynamical} \rspace  &  &  &  &  &  &  &  &  &  &  &  &  &  &  &  &  &  & \f &  &  &  &  & \rspace Nature \\

\rowcolor{rowbackground}
Tong,et.al.\cite{tong2010vulnerability} \rspace  &  &  &  &  &  &  &  &  &  & \f &  &  &  &  &  &  &  &  &   &  &  & \f &  \rspace ICDM \\

Tong,et.al.\cite{tong2012gelling} \rspace  &  &  &  &  &  &  &  &  &  &  & \f &  &  &  &  &  &  &  & \f & \f &  &  &  \rspace CIKM \\

\rowcolor{rowbackground}
Torres,et.al.\cite{torres2020node} \rspace  &  &  &  &  &  &  &  &  &  & \f &  &  &  &  &  &  &  & \f &  &  &  &  &  \rspace arXiv \\

Trajanovski,et.al.\cite{trajanovski2013robustness} \rspace  &  &  &  &  & \f &  &  &  & \f &  &  &  &  &  &  &  &  & \f &  &  &  &  & \rspace JOCN \\

\rowcolor{rowbackground}
Vespignani,et.al.\cite{vespignani2010fragility} \rspace  &  &  &  &  &  &  &  &  & \f &  &  &  &  &  &  &  &  &  &  &  &  &  & \rspace Nature \\

Wang,et.al.\cite{wang2008attack} \rspace  &  &  &  &  &  &  &  &  &  &  &  &  &  &  &  &  &  & \f &  &  &  &  &  \rspace Physica A \\

\rowcolor{rowbackground}
Wang,et.al.\cite{wang2014improving} \rspace  &  &  &  &  &  &  &  &  &  &  &  &  &  &  & \f &  & \f &  & \f & \f &  &  & \rspace EPJ B \\

Watts,et.al.\cite{watts1998collective} \rspace  &  &  &  &  & \f &  &  & \f & \f &  &  &  &  &  &  &  &  &  &  & \f &  &  & \rspace Nature \\

\rowcolor{rowbackground}
Wu,et.al.\cite{jun2010natural} \rspace  &  &  & \f &  &  &  &  &  &  &  &  & \f &  &  & \f &  &  &  & \f &  &  &  &  \rspace CPL \\

Wu,et.al.\cite{wu2011spectral} \rspace  &  &  & \f & \f & \f &  &  &  & \f &  &  & \f &  &  & \f &  &  & \f & \f &  &  &  & \rspace SMC \\

\rowcolor{rowbackground}
Xia,et.al.\cite{xia2010cascading} \rspace  &  &  &  &  &  & \f &  &  & \f &  &  &  &  &  &  &  &  & \f &  &  &  &  &  \rspace Physica A \\

Yang,et.al.\cite{yang2015improving} \rspace  &  &  &  &  &  & \f &  &  & \f &  &  &  &  &  &  &  &  & \f &  & \f &  &  & \rspace PLOS ONE \\

\rowcolor{rowbackground}
Yazdani,et.al.\cite{yazdani2010robustness} \rspace  &  & \f & \f &  &  & \f & \f &  &  &  & \f &  &  &  & \f &  &  & \f & \f &  &  &  &  \rspace CNSNS \\

Yazdani,et.al.\cite{yazdani2011complex} \rspace  &  & \f & \f & \f & \f & \f &  & \f &  &  & \f &  &  &  & \f &  &  & \f &  &  &  &  &  \rspace CHAOS \\

\rowcolor{rowbackground}
Zeng,et.al.\cite{zeng2012enhancing} \rspace  &  &  &  &  & \f &  & \f & \f &  &  &  &  &  &  &  &  &  & \f & \f &  &  &  & \rspace PRE \\

Zhao,et.al.\cite{zhao2004attack} \rspace  &  &  &  &  &  &  &  &  & \f &  &  &  &  &  &  &  &  & \f &  &  &  &  &  \rspace PRE \\

\rowcolor{rowbackground}
Zhao,et.al.\cite{zhao2014immunization} \rspace  &  &  &  &  &  &  &  &  &  &  &  &  &  &  &  &  &  &  &  &  &  & \f & \rspace PLOS ONE \\

\bottomrule
\end{tabular}

\caption{
Summary of works studied in this survey, each row is one work.
Columns are grouped into one of three categories---robustness measures, attacks and defenses---corresponding to primary paper sections (except ``where'').
In addition, we divide the robustness measure columns into three categories based on whether it uses the graph, adjacency matrix, or Laplacian matrix, from left to right, respectively (using dashed lines)
%\textcolor{metric-color}{robustness measures}
%\textcolor{attack-color}{attacks}
%\textcolor{defense-color}{defenses}
}
\label{table:papers}
\end{table*}

\begin{table*}[h]
\footnotesize
    \begin{tabular}[t]{l l}
    \toprule
    \textbf{Symbol} & \textbf{Graph Definitions} \\
    \midrule
        $G(\altmathcal{V}$, $\altmathcal{E})$ & graph $G$, set of nodes $\altmathcal{V}$, edges $\altmathcal{E}$ \\
        $n, m$ & number nodes $|V|$, edges $|\altmathcal{E}|$ \\
        $\kappa$, $\kappa_v$, $\kappa_e$ & binary, vertex, edge connectivity \\
        $\bar{d}$ & average geodesic distance  \\
        $d_{max}$ & graph diameter \\
        $\bar{b}_v$, $\bar{b}_e$ & avg. vertex, edge betweenness \\
        $C$ & global clustering coefficient \\
        $L$ & largest connected component \\
    \bottomrule
    \end{tabular}
    \hfill
    \begin{tabular}[t]{l l}
    \toprule
    \textbf{Symbol} & \textbf{Adjacency Matrix Definitions} \\
    \midrule
        $\bm{A}$ & adjacency matrix \\
        $\bm{A}_{i,j}$ & element at \textit{i}th row, \textit{j}th col. \\
        $\bm{u}(i)$ & eigenvector at position $i$ \\
        $\rho = \lambda_1$ & spectral radius = 1st eigenvalue \\
        $\lambda_d$ & spectral gap \\
        $\bar{\lambda}$ & natural connectivity \\
        %  $SC(i)$ & Sugraph centrality of node $i$ \\
        $\xi$ & spectral scaling \\
        $r_k$ & generalized robustness index \\
    \bottomrule
    \end{tabular}
    \hfill
    \begin{tabular}[t]{l l} 
    \toprule
    \textbf{Symbol} & \textbf{Laplacian Matrix Definitions} \\
    \midrule
        $\bm{L}$ & laplacian matrix \\
        $\bm{L}_{i,j}$ & element at \textit{i}th row, \textit{j}th col. \\
        $\bm{D}$ & diagonal matrix \\
        $d_i$ & degree of node $i$ \\
        $\mu_1$ & smallest eigenvalue of $\bm{L}$ \\
        $\mu_2$ & algebraic connectivity of $\bm{L}$ \\
        $T$ & number of spanning trees \\
        $R$ & effective resistance \\
    \bottomrule
    \end{tabular}
\caption{Symbols and Definition Tables. We divide symbols and definitions based on whether it corresponds to use with the graph, adjacency matrix or Laplacian matrix. From left to right, symbol and definition tables for the graph, adjacency matrix and Laplacian matrix.}
\label{table:notation}
\end{table*}

\subsection{Survey Organization.}

Figure~\ref{fig:crown} shows a visual overview of this survey's structure and Table~\ref{table:papers} summarizes representative works. 
The remainder of this paper is divided into seven parts. 
Each major component (measures, attacks and defenses) is given one or more sections, ordered to motivate how each component builds on top of the other. 

\begin{itemize}[leftmargin=0.87cm, itemsep=0.1cm, topsep=0.2cm]

    \item[
    \whattag{ \textbf{\textcolor{white}{$\pmb{\S}$ \ref{section:metrics}}} }
    ] \textbf{Summarizing \& Comparing Robustness Measures}\\
    We summarize recent and classical robustness measures, along with how each measure is \textit{linked} to the evaluation of graph vulnerability and robustness.
    We then create a table summarizing each robustness measure, allowing users to compare each measures in a simple manner.
    
    \item[
    \whytag{ \textbf{\textcolor{white}{$\pmb{\S}$ \ref{section:attack}}} }
    ] \textbf{Discussing Network Failures and Targeted Attacks}\\
    We discuss natural failures and targeted attacks strategies in relation to common graph topologies.
   
    \item[
    \whotag{ \textbf{\textcolor{white}{$\pmb{\S}$ \ref{section:defense}}} }
    ] \textbf{Analyzing Network Defense Mechanisms}\\
    We summarize common network defense mechanisms that are used to mitigate damage across a variety of network topologies and attacks.

\end{itemize}

Section~\ref{sec:future} presents research directions and open problems that we have gathered and distilled from the literature survey. 
Section~\ref{sec:conclusion} concludes the survey. 

\section{Robustness Measures}\label{section:metrics}
We begin by summarizing 18 recent and classic robustness measures, dividing each measure into one of three categories depending on whether it uses the graph (Section~\ref{subsection:robustness-graph}), adjacency matrix (Section~\ref{subsection:robustness-adj}) or Laplacian matrix (Section~\ref{subsection:robustness-lap}).
After describing each measure, we describe its \textit{link} to the study of network robustness. 
We make note of some additional robustness measures such as scattering number~\cite{jung1978class}, tenacity~\cite{cozzens1995tenacity}, integrity~\cite{barefoot1987integrity}, fault diameter~\cite{krishnamoorthy1987fault}, toughness~\cite{chvatal1973tough} and isoperimetric number~\cite{mohar1989isoperimetric}.
However, we do not consider them for evaluation since they are combinatorial measures for general graphs~\cite{chan2014make}.

\subsection{Measures Based on Graph Connectivity}\label{subsection:robustness-graph}
We review $9$ common graph robustness measures, each of which takes as input an undirected, unweighted graph $G=(\altmathcal{V},\altmathcal{E})$, where $\altmathcal{V}$ is the set of vertices, $\altmathcal{E}\subseteq \altmathcal{V} \times \altmathcal{V}$ is the set of edges.
We let $n = |\altmathcal{V}|$ and $m = |\altmathcal{E}|$ as the number of vertices and number of edges, respectively.

\subsubsection{Binary Connectivity ($\kappa$)}\label{subsec:binary-connectivity}
A classical graph measure which determines whether or not a graph is connected ($\kappa$=1) or unconnected ($\kappa$=0) by examining whether all pairs of vertices have a connecting path. 
A graph is considered unconnected if at least one pair of vertices does not have a connecting path.
This can be calculated using breadth-first or depth-first search, starting at any vertex with time complexity $O(n+m)$.

\vspace{0.25cm}
\noindent\textbf{Robustness link.} Practically speaking, binary connectivity is a poor measure of network robustness since it only identifies whether a network is disconnected.

\subsubsection{Vertex Connectivity ($\kappa_v$)}\label{subsec:vertex-connectivity}
An extension of binary connectivity, vertex connectivity $\kappa_v(G)$ is the minimal number of vertices that need to be removed to disconnect the graph $\kappa_v(G)$ = $min\{ \kappa_v(u, v) \mid \text{ unordered pair } (u, v) \in G \}$.
Assume $(u,v)\not\in E$, then $\kappa_v(u, v)$ is the minimal number of vertices that would destroy every path between vertices $u$ and $v$.
If $(u,v)\in E$, then $\kappa_v(u, v) = n - 1$~\cite{henzinger2000computing}.
Computing $\kappa_v(G)$ can be reduced to a max-flow problem with a time complexity of $O(n^{8/3}m)$~\cite{matula1987determining}.

\vspace{0.25cm}
\noindent\textbf{Robustness link.} This measure has a natural relationship to the robustness of the graph, since $\kappa_v(G)$ increases as the graph becomes harder to disconnect.

\subsubsection{Edge Connectivity ($\kappa_e$)}\label{subsec:edge-connectivity}
Also an extension of binary connectivity, edge connectivity $\kappa_e(G)$ is the minimal number of edges that can be removed to disconnect the graph $\kappa_e(G)$ = $min\{\kappa_e(u, v) \mid \text{ unordered pair } (u, v) \in G$\}. Let $u$, $v$, be a pair of distinct vertices in graph $G$, $\kappa_e(u, v)$ is the minimal number of deleted edges that would disconnect all paths between $u$ and $v$. Calculating $\kappa_e(u, v)$ for each unordered pair, we can ascertain the minimal edge connectivity.
The best known algorithm has a time complexity of $O(nm)$~\cite{matula1987determining,esfahanian2013connectivity}.
For an incomplete graph, an interesting property of vertex and edge connectivity is that $\kappa_v(G) \leq \kappa_e(G) \leq \delta_{min}(G)$, where $\delta_{min}(G) = min\{deg(v) \mid v \in G\}$~\cite{whitney1992congruent}.

\vspace{0.25cm}
\noindent\textbf{Robustness link.} 
Similar to $\kappa_v(G)$, this measure naturally ties to graph robustness, since $\kappa_e(G)$ increases as the graph becomes harder to disconnect.

\subsubsection{Diameter ($d$)}\label{subsec:diameter}
The diameter $d$ of a connected graph is the longest shortest path between all pairs of nodes $d(G)$ = $max\{d(u, v) \mid \text{ unordered pair } (u, v) \in G\}$, where $d(u, v)$ is the shortest path between vertices $u$ and $v$.
In order to calculate the diameter of a network all pairs of shortest paths need to be computed, requiring a time of complexity of $O(n^3)$ using the Floyd–Warshall algorithm~\cite{warshall1962algorithm}.

\vspace{0.25cm}
\noindent\textbf{Robustness link.}
The diameter has an intuitive connection to robustness, where a decreasing diameter implies better robustness.
This is because as the diameter of the network shrinks (i.e., through edge addition),  the longest shortest paths between distant vertices are reduced, and the network becomes more tightly coupled.

\subsubsection{Average Distance ($\bar{d}$)}\label{subsec:average-distance}
The average geodesic distance $\bar{d}$ in Equation~\ref{eq:avg_dist_1} provides a measure of network connectivity by calculating the average distance between all pairs of nodes in the graph.

\begin{equation}\label{eq:avg_dist_1}
    \bar{d} = \frac{2}{n(n-1)}\sum_{u\in V}\sum_{\substack{v\in V \\ u\neq v}}d(u, v)
\end{equation}

\noindent For each unordered pair of nodes $(u, v)\in G$ in the graph, the shortest path $d(u, v)$ is computed using the Floyd-Warshall algorithm~\cite{warshall1962algorithm}, then summed and normalized by $\frac{2}{n(n-1)}$ to account for bi-directional paths $d(u, v) = d(v, u)$ in undirected graphs.
This measure is commonly modified to account for disconnected graphs by computing the average inverse distance, also known as the \textit{efficiency}, as shown in Equation~\ref{eq:avg_dist_2}.
Unfortunately, both average distance and efficiency have a time complexity of $O(n^3)$ due to the all pairs shortest path computation. 

\begin{equation}\label{eq:avg_dist_2}
    \bar{d} = \frac{2}{n(n-1)}\sum_{u\in V}\sum_{\substack{v\in V \\ u\neq v}}\frac{1}{d(u, v)}
\end{equation}

\vspace{0.25cm}
\noindent\textbf{Robustness link.}
This measure has a close relationship to network connectivity, where a smaller average distance implies a more robust graph, since we are reducing the average distance between any pair of nodes in the network.
In contrast, a larger efficiency implies a more robust graph due to inverse computation.

\subsubsection{Average Vertex Betweenness ($\bar{b}_v$)}\label{subsec:vertex-betweenness}
The average vertex betweenness $\bar{b}_v$ in Equation~\ref{eq:avg-vertex-betweenness} is the summation of vertex betweenness $b_u$ for every node $u\in V$,

\begin{equation}\label{eq:avg-vertex-betweenness}
    \bar{b}_v = \sum_{u\in V} b_u
\end{equation}

\noindent where vertex betweenness $b_u$ for node $u$ is defined in Equation~\ref{eq:vertex-betweenness} as the number of shortest paths that pass through $u$ out of the total possible shortest paths.

\begin{equation}\label{eq:vertex-betweenness}
    b_u = \sum_{s\in V}\sum_{\substack{t\in V \\ s\neq t\neq u}} \frac{n_{s,t}(u)}{n_{s,t}}
\end{equation}

\noindent Here $n_{s, t}(u)$ is the number of shortest paths between $s$ and $t$ that pass through $u$ and $n_{s, t}$ is the total number of shortest paths between $s$ and $t$~\cite{freeman1977set}.
Interestingly, the average vertex betweenness $\bar{b}_v$ can be represented as a linear function of the average distance $\bar{d}$~\cite{ellens2013graph}, as shown in Equation~\ref{eq:average-distance-vertex-betweenness}, implying a strong connection between the robustness properties of average vertex betweenness and average distance.

\begin{equation}\label{eq:average-distance-vertex-betweenness}
    \bar{b}_v = \frac{1}{2}(n-1)(\bar{d}+1)
\end{equation}

\noindent Computing the vertex betweenness centrality $b_u$ for a single node $u$ has a time complexity of $O(nm)$ using Brandes' algorithm~\cite{brandes2001faster}.
Therefore, the average vertex betweenness $\bar{b}_v$ has a time complexity of $O(n^2m)$.

\vspace{0.25cm}
\noindent\textbf{Robustness link.}
Average vertex betweenness has a natural connection to graph robustness since it measures average network throughput of the vertices.
The smaller the average the more robust the network, since the shortest paths are more evenly distributed across each node, rather than relying on a few central nodes.

\subsubsection{Average Edge Betweenness ($\bar{b}_e$)}\label{subsec:edge-betweenness}
Similar to vertex betweenness, edge betweenness is defined as the number of shortest paths that pass through an edge $e$ out of the total possible shortest paths.
Since the formula and intuition for calculating average edge betweenness is nearly identical to average node betweenness, we define it Equation~\ref{subsec:vertex-betweenness} below and refer the reader to Section~\ref{subsec:vertex-betweenness} for additional detail.

\begin{equation}\label{eq:graph-edge-betweenness}
    \bar{b}_e = \sum_{e\in E}\sum_{s\in V}\sum_{\substack{t\in V \\ s\neq t}}\frac{n_{s,t}(e)}{n_{s,t}}
\end{equation}

Similar to $\bar{b}_v$, average edge betweenness can be represented as a linear function of the average distance $\bar{d}$,

\begin{equation}\label{eq:average-distance-edge-betweenness}
    \bar{b}_e = \frac{n(n-1)}{2m}\bar{d}
\end{equation}

\noindent implying that some of the robustness properties for $\bar{d}$ will hold for $\bar{b}_e$ as well.
The time complexity for average edge betweenness is also $O(n^2m)$.

\vspace{0.25cm}
\noindent\textbf{Robustness link.}
Average edge betweenness has similar robustness properties to average vertex betweenness.
The smaller the average the more robust the network, since the shortest paths are more evenly distributed across each edge, rather than relying on a few central edges. 

\subsubsection{Global Clustering Coefficient ($C$)}\label{subsec:gcc}
The global clustering coefficient $C$ is based on the number of triplets of nodes in the graph, and provides an indication of how well nodes tend to cluster together.
By definition, a triplet is three nodes connected by either two edges (open triplet) or three edges (closed triplet); where a closed triplet is called a triangle. 
Each triangle contains three closed triplets, one centered on each node.
In order to measure the global clustering coefficient, we count the number of closed triplets (or 3x the number of triangles) and divide it by the total number of both closed and open triplets, as in Equation~\ref{eq:gcc}.

\begin{equation}\label{eq:gcc}
    C = \frac{\text{closed triplets}}{\text{closed triplets} + \text{open triplets}}
\end{equation}

Alternatively, we can view the global clustering coefficient as the average possible fraction of interconnections among each node $v\in G$ as in Equation~\ref{eq:gcc_2}

\begin{equation}\label{eq:gcc_2}
    C = \frac{1}{n} \sum_{n\in V} \frac{2\cdot N_v}{\delta_v(\delta_v - 1)}
\end{equation}

\noindent where $N_v$ is the number of edges between neighbors of $v$, and $\delta_v$ is the degree of node $v$. 
The time complexity for computing the global clustering coefficient is $O(n\cdot d^2_{max})$, where $d^2_{max}$ is the size of the largest adjacency list across all vertices in the graph~\cite{green2013faster}.

\vspace{0.25cm}
\noindent\textbf{Robustness link.}
As the global clustering coefficient increases, we begin to see the formation of tight knit groups (due to an increased density of triangles)---which creates redundant pathways between neighbors, and increases the overall robustness of the network.

\subsubsection{Largest Connected Component ($L$)}\label{subsec:lcc}
This measure provides an indication of a graph's connectivity by measuring the fraction of nodes contained in the largest connected component.
This is calculated by determining the maximal set of vertices $S \subset V$ such that for each $u\in S$ and $v\in S$, there exits a path from $u$ to $v$ in $G$. 
Intuitively, $L$ provides a measure of network availability i.e., what percentage of the nodes can be reached, and is measured as shown in Equation~\ref{eq:lcc}.

\begin{equation}\label{eq:lcc}
    L = \frac{|\{d(u, v) \neq \infty \mid \text{unordered pair} (u, v)\in G \}|}{n} 
\end{equation}

\noindent The time complexity to find the largest connected component is $O(n + m)$, since each BFS (or DFS) call takes linear time in the number of edges and vertices for each component; and since each component is only searched once, all searches are linear in the number of edges and vertices.

\vspace{0.25cm}
\noindent\textbf{Robustness link.}
This is useful as a measurement of robustness since as the number of removed nodes and edges increases, there reaches a critical fraction where the network connectivity begins to collapse; which can be measured through the size of the largest connected component.

\subsection{Measures Based on Adjacency Matrix Spectrum}\label{subsection:robustness-adj}
The adjacency matrix $\bm{A}$ is a common network representation, often used when enumerate walks~\cite{butler2008eigenvalues}.
Formally, we say that $\bm{A}(G)$ is the adjacency matrix of $G$, where $\bm{A}\in \{0, 1\}^{n\times n}$ and $\bm{A}_{i,j} = \bm{A}_{j,i}$ = $1$ if vertex $v_i$ and $v_j$ are adjacent and $\bm{A}_{i,j} = \bm{A}_{j,i}$ = $0$ otherwise. This can be seen in Equation~\ref{eq:adj-matrix}.

\begin{equation}\label{eq:adj-matrix}
\small
\begin{aligned}
    \bm{A}_{ij} = 
    \begin{cases}
        1, & \text{if $i$ is adjacency to $j$} \\
        0,              & \text{otherwise}
    \end{cases}    
\end{aligned}
\end{equation}

It follows that $\bm{A}(G)$ is a real symmetric matrix with real eigenvalues $\lambda_1 \geq \lambda_2 \geq...\geq \lambda_n$. The set of eigenvalues $\{\lambda_1, \lambda_2,...\lambda_n\}$ is called the spectrum of $\bm{A}$, with corresponding eigenvectors $\bm{u}_1, \bm{u}_2,...,\bm{u}_n$.
Several robustness measures have been based on the spectrum of the adjacency matrix; we address 5 of the most common ones below.

\subsubsection{Spectral Radius ($\rho$)}\label{subsec:spectral-radius}
The largest eigenvalue $\lambda_1$ of an adjacency matrix $\bm{A}$ is called the spectral radius $\rho$.
The spectral radius is closely related to the \textit{path capacity} or \textit{loop capacity} of the graph. 
That is, the number of walks of length $k$ ($k=2, 3, 4...$) gives an indication of how well connected the graph is. 
If the graph has many loops and paths, then the graph is well connected i.e., larger $\lambda_1$~\cite{tong2010vulnerability,butler2008eigenvalues,mieghem2011decreasing}.
It has been shown in \cite{wang2003epidemic} that the spectral radius is also closely tied to the epidemic threshold $\tau$ of a network in the flu-like SIS (susceptible-infected-susceptible) model.
In particular, they prove that $\frac{\beta}{\delta} < \tau=\frac{1}{\lambda_{1}}$, where $\beta$ is the birth rate of a virus or disease and $\delta$ is the cure rate. 
This means for a given virus strength, an epidemic is more likely to occur on a graph with larger $\lambda_1$.
While this seems contradictory at first glance, increased vulnerability to virus propagation actually implies a graph is more robust to natural failures and targeted attacks. 
The time complexity for computing the spectral radius is $O(m)$, since computing the first eigenpair in sparse matrix form is linear with respect to the number of edges~\cite{chen2015node}.

\vspace{0.25cm}
\noindent\textbf{Robustness link.}
As a robustness measure, a larger $\lambda_1$ indicates a more robust graph to random failures and attack, along with increased susceptibility to virus propagation~\cite{le2015met,tong2010vulnerability,chen2015node,saha2015approximation,tong2012gelling,chan2016optimizing}. 
This is because the backup paths and redundancy in a network are the same mechanisms that allow a virus to quickly propagate.

\subsubsection{Spectral Gap ($\lambda_{d}$)}\label{subsec:spectral-gap}
The difference between the largest and second largest eigenvalues of the adjacency matrix ($\lambda_1 - \lambda_2$) is called the spectral gap $\lambda_d$.
As a robustness measure, the spectral gap is a simple way to estimate the robustness of a graph---if the spectral gap is large, the graph shows good robustness; if it is small, the robustness is poor~\cite{chan2016optimizing,malliaros2012fast}.
This is because a large spectral gap is indicative of a network with good expansibility properties, while a small spectral gap indicates a network with bottlenecks and bridges~\cite{estrada2006network}.
This ability to communicate the existence of bridges in the graph is a unique property of the spectral gap, and not found in the spectral radius.
The time complexity to compute the spectral gap is $O(m + n)$~\cite{chen2015fast}.

\vspace{0.25cm}
\noindent\textbf{Robustness link.}
A unique link between the spectral gap and network robustness is the ability to identify bridges and bottlenecks in the graph~\cite{estrada2006network}. 
Unfortunately, it is not evident how large the spectral gap needs to be for a graph to be characterized as robust~\cite{malliaros2012fast}.

\subsubsection{Natural Connectivity ($\bar{\lambda}$)}\label{subsec:natural-connectivity}
Natural connectivity can be interpreted as the ``average eigenvalue'' of the adjacency matrix $\bar{\lambda}$~\cite{jun2010natural}, and is defined in Equation~\ref{eq:natural_connectivity}.

\begin{equation}\label{eq:natural_connectivity}
    \bar{\lambda}(G) = ln(\frac{1}{n}\sum_{i=1}^{n}e^{\lambda_i})
\end{equation}

\noindent Natural connectivity has a physical and structural interpretation that is tied to the connectivity properties of a network.
It is often used to measure the availability of alternative pathways in a network, through the weighted number of closed walks~\cite{chan2014make}. 
A walk of length $k$ in a graph is defined as an alternating sequence of vertices and edges $v_0e_1v_1e_2...e_kv_k$, where $v_i\in V$ and $e_i = (v_{i-1}, v_i)\in E$.
A walk is said to be closed if $v_0 = v_k$. 
Closed walks are closely related to the \textit{natural connectivity} and \textit{subgraph centrality} ($SC$), which we can relate through Equation~\ref{eq:sc}.

\begin{equation}\label{eq:sc}
    SC(G) = \sum_{i=i}^{n} SC(i) = \sum_{i=1}^{n}\sum_{k=0}^{\inf} \frac{(\bm{A}^k)_{ii}}{k!} = \sum_{i=1}^{n}\sum_{k=0}^{\inf} \frac{\lambda_i^k}{k!} = \sum_{i=1}^{n} e^{\lambda_i}
\end{equation}

\noindent Here $(\bm{A}^k)_{ii}$ is the number of closed walks of length $k$ on node $i$, where $k!$ scales the weighted sum so that it does not diverge and longer walks are counted less~\cite{estrada2005subgraph,jun2010natural,chan2014make}. 
We further explore the relationship between subgraph centrality and robustness in Sections~\ref{subsec:spectral-scaling} and~\ref{subsec:robustness-index}.
For now, we rewrite Equation~\ref{eq:sc} in relation to natural connectivity as shown in Equation~\ref{eq:sc2}.

\begin{equation}\label{eq:sc2}
    \bar{\lambda}(G) = ln(\frac{1}{n}\sum_{i=1}^{n}e^{\lambda_i}) = ln(\frac{1}{n}SC(G))
\end{equation}

\noindent Natural connectivity has a time complexity of $O(n^3)$ since it computes the full spectrum of the adjacency matrix~\cite{nardo2018applications}.

\vspace{0.25cm}
\noindent\textbf{Robustness link.}
Natural connectivity has a close relation to network topology and dynamical processes on the graph~\cite{chan2014make}.
This can be seen by its close connection to the number of closed walks, which naturally captures the notion of network connectivity and alternative pathways in a network.
A larger $\bar{\lambda}$ indicates a more robust graph.

\subsubsection{Spectral Scaling ($\xi$)}\label{subsec:spectral-scaling}
When a network is both sparse and highly connected it is said to have ``good expansion'' (GE), also known as an expander graph or a ``good expansion network'' (GEN)~\cite{hoory2006expander,estrada2006network}. 
Intuitively, we can think of an expander graph as a network lacking bridges or bottlenecks, making it hard to disconnect through the removal of a few nodes or edges. 
GE is closely related to the spectral gap $\lambda_d$ of a network, and can be used to identify the GE property if $\lambda_d$ is sufficiently large (i.e., $\lambda_1 \gg \lambda_2$)~\cite{estrada2006network}.
The question is, how large should the spectral gap be in order for a network to be considered a GEN?

Estrada~\cite{estrada2006network} proposes to solve this problem by combining the \textit{spectral gap} with \textit{subgraph centrality} ($SC$).
In particular, they propose to use odd subgraph centrality $SC_{odd}$ to measure the number of odd length closed walks that a node participates in.
This helps to avoid trivial closed walks (i.e., paths) occurring from even length closed walks $SC_{even}$.
Rewriting Equation~\ref{eq:sc}, we can view $SC$ in terms of it's even and odd components~\cite{estrada2005spectral}:

\begin{equation}\label{eq:sc_components}
\begin{split}
    SC(i) = \sum_{j=1}^n \bm{u}_{j}(i)^2 cosh(\lambda_j) + & \sum_{j=1}^n \bm{u}_{j}(i)^2 sinh(\lambda_j) \\ = SC_{even}(i) + & SC_{odd}(i)
\end{split}
\end{equation}

\noindent Expanding $SC_{odd}$ into two components based on the first and subsequent eigenvalues:

\begin{equation}\label{eq:sc_odd_expanded}
    SC_{odd}(i) = [\bm{u}_1(i)]^2 sinh(\lambda_1) + \sum_{j=2}[\bm{u}_j(i)]^2 sinh(\lambda_j)
\end{equation}

\noindent Since we know that networks with good expansibility have $\lambda_1 \gg \lambda_2$, we can say that:

\begin{equation}\label{eq:sc_transition}
    [\bm{u}(i)]^2 sinh(\lambda_1) \gg \sum_{j=2}[\bm{u}_j(i)]^2 sinh(\lambda_j)
\end{equation}

\noindent and therefore rewrite Equation~\ref{eq:sc_odd_expanded} using the inequality from Equation~\ref{eq:sc_transition} as follows:

\begin{equation}\label{eq:sc_approx}
    SC_{odd}(i) \approx [\bm{u}_1(i)]^2 sinh(\lambda_1)
\end{equation}

\noindent From Equation~\ref{eq:sc_approx} we observe that the subgraph centrality is proportional to the first eigenvector $\bm{u}_1$, yielding the following equation:

\begin{equation}\label{eq:sc_proportional}
    \bm{u}_1(i) \propto \bm{A}[SC_{odd}(i)]^\eta
\end{equation}

\noindent where $\bm{A}$ = $[sinh(\lambda_1)]^{-0.5}$ and $\eta \approx 0.5$ This implies a linear correlation between $\bm{u}_1(i)$ and $SC_{odd}(i)$ for networks with good expansion. 
In a log-log scale Equation~\ref{eq:sc_proportional} can be rewritten as:

\begin{equation}
    log[\bm{u}_i(i)] = log\bm{A} + \eta log[SC_{odd}(i)]
\end{equation}

\noindent As a result, a log-log plot of $\bm{u}_{1}(i)$ vs. $SC_{odd}(i)$ shows a linear fit with slope $\eta \approx 0.5$ with an intercept of $log[\bm{A}]$ if the network has GE.
In order to quantify this into a robustness measure, \cite{estrada2006network} proposes the formula for \textit{spectral scaling} in Equation~\ref{eq:expansion_character}.

\begin{equation}\label{eq:expansion_character}
    \xi(G) = \sqrt{\frac{1}{n} \sum_{i=1}^{n} \{log[\bm{u}_1(i)] - [log\bm{A} + \eta log[SC_{odd}(i)]] \}^2 }
\end{equation}

\noindent We note that spectral scaling calculates  $SC_{odd}(i)$ using the full spectrum of the adjacency matrix, not just the first eigenpair.
As such, the time complexity is $O(n^3)$ due to the computation of the full adjacency matrix spectrum~\cite{wu2018scalable}.

\vspace{0.25cm}
\noindent\textbf{Robustness link.}
As a robustness measure, the closer the value of $\xi$ to zero, the better the expansion properties and the more robust the network.
Formally, a network is considered to have GE if $\xi < 10^{-2}$, the correlation coefficient $r < 0.999$ and the slope is $0.5$.
While this method improves upon the spectral gap, it still suffers from a few shortcomings, including---(i) not scalable to large graphs due to computing the full adjacency matrix spectrum; and (ii) not applicable to bipartite graphs which do not contain odd length closed walks.

\subsubsection{Generalized Robustness Index ($r_k$)}\label{subsec:robustness-index}
This measure is a fast approximation of \textit{spectral scaling}, which includes only the top $k$ eigenpairs in the subgraph centrality calculation, since only the first few eigenvalues are large, with most of the eigenvalues symmetric around zero. 
In many real world graphs $k \ll n$, and therefore, $k$ can be considered a low-rank approximation of the adjacency matrix $\bm{A}$~\cite{malliaros2012fast}.
This allows us to compute only a few eigenpairs of $\bm{A}$ while capturing a majority of the spectrum information; making the measure scalable to large graphs.
This process can be seen in Equation~\ref{eq:nsc}, where the modified subgraph centrality is referred to as \textit{normalized subgraph centrality} $(NSC)$.

\begin{equation}\label{eq:nsc}
    NSC_k(i) = \sum_{j=1}^{k} sinh(\lambda_j), \forall i\in \altmathcal{V}
\end{equation}

\noindent Based on the normalized subgraph centrality $NSC_k$, \cite{malliaros2012fast} proposes the \textit{generalized robustness index} $r_k$:

\begin{equation}
    r_k = \sqrt{\frac{1}{n} \sum_{i=1}^{n}  \{log[\bm{u}_1(i)] - [ log\bm{A} + \frac{1}{2}log[NSC_k(i)] ]\}^{2} }
\end{equation}

\noindent where $\bm{A}$ = $[sinh(\lambda_1)]^{-0.5}$. 
In practice, $k$ can be extremely small while still achieving high accuracy ($k\leq 30$)~\cite{malliaros2012fast}.
This reduces the time complexity of spectral scaling to either $O(kn^2)$\cite{wu2018scalable} or $(mk + nk^2)$~\cite{chen2015fast} depending on the implementation.

\vspace{0.25cm}
\noindent\textbf{Robustness link.}
As a robustness measure, a smaller $r_k$ implies better robustness.
Asides from the computational benefit of the low rank approximation, this measure is very similar to spectral scaling. We refer the reader to Section~\ref{subsec:spectral-scaling} for additional information.

\begin{table*}[t]
\centering

\setlength{\tabcolsep}{15pt}
\renewcommand\arraystretch{1.5}

 \begin{tabular}{lll@{\hspace{5mm}}l@{\hspace{5mm}}l@{\hspace{5mm}}l@{\hspace{5mm}}l} 
 
 \toprule
    
  \textbf{Robustness Measure} & \thead[l]{Category} & \multicolumn{5}{l}{\textbf{Application to Network Robustness}} \\
  
 \midrule

 Vertex connectivity & graph  &  higher value & $\Rightarrow$ & harder to disconnect graph & $\Rightarrow$ & higher robustness \\
 
  \rowcolor{rowbackground}
 Edge connectivity & graph & higher value & $\Rightarrow$ & harder to disconnect graph & $\Rightarrow$ & higher robustness \\

 Diameter & graph & lower value & $\Rightarrow$ & stronger connectivity & $\Rightarrow$ & higher robustness \\
 
  \rowcolor{rowbackground}
 Average distance & graph & lower value & $\Rightarrow$ & stronger connectivity & $\Rightarrow$ & higher robustness \\
 
 Average inverse distance & graph & higher value & $\Rightarrow$ & stronger connectivity & $\Rightarrow$ & higher robustness  \\
 
 \rowcolor{rowbackground}
 Average vertex betweenness & graph & lower value  & $\Rightarrow$ & more evenly distributed load  & $\Rightarrow$ & higher robustness \\
 
 Average edge betweenness & graph &  lower value  & $\Rightarrow$ & more evenly distributed load  & $\Rightarrow$ & higher robustness \\
 
 \rowcolor{rowbackground}
 Global clustering coefficient & graph & higher value & $\Rightarrow$ & more triangles & $\Rightarrow$ & higher robustness  \\
 
 Largest connected component & graph & higher value & $\Rightarrow$ & better connected graph & $\Rightarrow$ & higher robustness \\
 
 \midrule
 
 \rowcolor{rowbackground}
 Spectral radius & adjacency & larger value & $\Rightarrow$ & stronger connectivity & $\Rightarrow$ & higher robustness \\
 
 Spectral gap & adjacency & higher value & $\Rightarrow$ & fewer bottlenecks & $\Rightarrow$ & higher robustness \\
 
 \rowcolor{rowbackground}
 Natural connectivity & adjacency & higher value & $\Rightarrow$ & more alternative pathways & $\Rightarrow$ & higher robustness \\
 
 Spectral scaling & adjacency & lower value & $\Rightarrow$ & fewer bottlenecks & $\Rightarrow$ & higher robustness \\
 
 \rowcolor{rowbackground}
 Generalized robustness index & adjacency & lower value & $\Rightarrow$ & fewer bottlenecks & $\Rightarrow$ & higher robustness \\
 
 \midrule
 
 Algebraic connectivity & laplacian & higher value & $\Rightarrow$ & harder to disconnect & $\Rightarrow$ & higher robustness \\
 
 \rowcolor{rowbackground}
 Number of spanning trees & laplacian & higher value & $\Rightarrow$ & more alternative pathways & $\Rightarrow$ & higher robustness \\
 
 Effective resistance & laplacian & lower value & $\Rightarrow$ & more alternative pathways & $\Rightarrow$ & higher robustness \\
 
\bottomrule
\end{tabular}
\caption{Comparison of robustness measures.
Measures are grouped based on whether they use the \textit{graph}, \textit{adjacency} or \textit{Laplacian} matrix.
For each measure, we briefly describe it's application to measuring network robustness.
}
\label{table:metric_comparison}
\end{table*}

\subsection{Measures Based on Laplacian Matrix Spectrum}\label{subsection:robustness-lap}

The Laplacian matrix $\bm{L}$ is often used when a problem can be related to spanning trees or the incidence of vertices and edges~\cite{butler2008eigenvalues}.
Formally, we say that $\bm{L}(G)$ is the Laplacian matrix of $G$, where $\bm{L} = \bm{D} - \bm{A}$ and $\bm{D}$ is the diagonal matrix with the degree on the diagonals. 
It follows that $\bm{L}_{i,j}$ = $d_i$ if $i = j$; $\bm{L}_{i,j}$ = $-1$ if $i$ is adjacent to $j$; and $\bm{L}_{i,j}$ = $0$ otherwise; where $d_i$ is the degree of node $i$. 
This can be seen in Equation~\ref{eq:laplacian-matrix}.

\begin{equation}\label{eq:laplacian-matrix}
\small
\begin{aligned}
    \bm{L}_{ij} = 
    \begin{cases}
        d_i, & \text{if } i=j \\
        -1, & \text{if $i$ is adjacency to $j$} \\
        0,              & \text{otherwise}
    \end{cases}    
\end{aligned}
\end{equation}

\noindent Since $\bm{D}$ and $\bm{A}$ are both symmetric and have real eigenvalues and an orthonormal basis of eigenvectors, the Laplacian matrix is positive semi-definite with nonnegative eigenvalues; with the smallest eigenvalue always being 0.
Hence we order the eigenvalues as follows $0$ = $\mu_1 \leq \mu_2 \leq...\leq \mu_n$, where the set of eigenvalues $\{\mu_1, \mu_2,...\mu_n\}$ is called the spectrum of $\bm{L}$.
Several robustness measures have been based on the spectra of the Laplacian matrix; we address 3 of the most prominent ones below.

\subsubsection{Algebraic Connectivity ($\mu_2$)}\label{subsec:algebraic-connectivity}
The second smallest eigenvalue of the Laplacian is called the algebraic connectivity $\mu_2$, also known as the Fiedler vector~\cite{fiedler1973algebraic}.
Since the Laplacian is symmetric, positive semi-definite and the rows sum up to 0, the eigenvalues are real and non-negative, with the smallest eigenvalue being zero; and the multiplicity of the zero eigenvalue related to the number of connected components.
For a disconnected graph, this means the algebraic connectivity is always zero.
The time complexity to compute the algebraic connectivity is $O(m+n)$~\cite{lehoucq1998arpack}

\vspace{0.25cm}
\noindent\textbf{Robustness link.}
The larger the algebraic connectivity $\mu_2$ the more robust the graph. This can be understood from its relationship to the characteristic path length of a network; and from its connection to node connectivity $\kappa_v$ and edge connectivity $\kappa_e$ of a graph, where $\mu_2$ serves as a lower bound $0 \leq \mu_2 \leq \kappa_v \leq \kappa_e \leq d_{min}$.
This means that a network with larger algebraic connectivity is harder to disconnect (i.e., more edges, nodes need to be removed)~\cite{chan2016optimizing}.

\subsubsection{Number of Spanning Trees ($T$)}\label{subsec:spanning-trees}
A spanning tree is a subgraph of $G$ containing $n$ nodes, $n$-$1$ edges and no cycles. 
We can visualize this as a graph connecting all the vertices with the minimum possible number of edges.
The number of spanning trees $T$ is the number of unique spanning trees that can be found in a graph. 
This measure was originally suggested as an indicator of a graph's ability to stay connected~\cite{weichenberg2004high}; where
Baras and Hovareshti expanded upon it as an indicator of network robustness~\cite{baras2009efficient}.
As a consequence of the Kirchoff's matrix-tree theorem~\cite{buekenhout1998number}, the number of spanning trees $T$ can be written as a function of the Laplacian eigenvalues as shown in Equation~\ref{eq:trees}.

\begin{equation}\label{eq:trees}
    T = \frac{1}{n} \prod_{i=2}^{n}\mu_i
\end{equation}

\noindent The time complexity for this measure is $O(n^3)$ due to the computation of the full Laplacian spectrum~\cite{tsironis2013accurate}.

\vspace{0.25cm}
\noindent\textbf{Robustness link.} The larger the number of spanning trees the more robust the graph. This can be viewed from the perspective of network connectivity, where a larger set of spanning trees provides a measure of alternative pathways.
Unfortunately this measure does not work for disconnected graphs since a spanning tree must include all vertices by definition.

\subsubsection{Effective Resistance ($R$)}\label{subsec:effective-resistance}
This measure views a graph as an electrical circuit where an edge $(i, j)$ corresponds to a resister of $r_{ij}$ = $1$ Ohm and a node $i$ corresponds to a junction.
As such, the effective resistance between two vertices $i$ and $j$, denoted $R_{ij}$, is the electrical resistance measured across nodes $i$ and $j$ when calculated using Kirchoff's circuit laws.
Extending this measure to the whole graph, we say the \textit{effective graph resistance} $R$ is the sum of resistances for all distinct pairs of vertices~\cite{ellens2013graph,ghosh2008minimizing}.
Klein and Randic proved this can be can be calculated based on the sum of the inverse non-zero Laplacian eigenvalues~\cite{klein1993resistance} as shown in Equation~\ref{eq:er}

\begin{equation}\label{eq:er}
    R = \frac{1}{2}\sum_{i, j}^{n} R_{ij} = n\sum_{i=2}^{n} \frac{1}{\mu_i}
\end{equation}

\noindent In addition, it has been shown that the effective resistance can be bounded by the algebraic connectivity~\cite{ellens2011effective} as shown in Equation~\ref{eq:er2}

\begin{equation}\label{eq:er2}
    \frac{n}{\mu_2} < R \leq \frac{n(n-1)}{\mu_2}
\end{equation}

\noindent The effective resistance has time complexity $O(n^3)$ due to computation of the full Laplacian spectrum.

\vspace{0.25cm}
\noindent\textbf{Robustness link.}
As a robustness measure, effective resistance measures how well connected a network is, where a smaller value indicates a more robust network~\cite{ghosh2008minimizing,ellens2013graph}. 
In addition, the effective resistance has many desirable properties, including the fact that it strictly decreases when adding edges~\cite{ellens2011effective}, and takes into account both the number of paths between node pairs and their length.

\subsection{Comparing Robustness Measures}
In Table~\ref{table:metric_comparison}, we highlight each robustness measure, the category it belongs to (graph, adjacency, Laplacian), and its application to network robustness.
By distilling all of this measure information into a single table, users can easily compare robustness measures.
This greatly assists in selecting robustness measures across domain specific criteria, where the user may have an idea of what the robustness measure needs to incorporate.
For example, users working with critical infrastructure systems wants to select a robustness measure that takes into account backup pathways, this information is now readily available, along with how to interpret the measure (e.g., lower is better).

\subsection{Practical Applications of Robustness Measures}
In Table~\ref{tab:measure-applications}, we organize a number of key robustness works studied across five high impact domains.
By summarizing this information into a single table, readers can quickly access relevant robustness research in the domain of their choice, simplifying the process of determining if a work is relevant to their application.
Despite the prevalence of robustness research, measuring the impact and adoption rate across industry and government organizations is challenging for multiple reasons:
(1) network telemetry used to create the graph is often considered sensitive material, and withheld from the public for security and privacy concerns;
(2) open-sourcing robustness related research can inadvertently provide malicious adversaries knowledge to reverse engineer defense systems; 
and (3) proprietary advances in robustness research can be considered trade secrets to prevent adoption by rival companies and government organizations.
As a result of the above challenges, our goal through Table~\ref{tab:measure-applications} is to focus on organizing works that study practical applications of network robustness using real-world data.

\begin{table}[t]
    \setlength{\tabcolsep}{10pt}
    \renewcommand{\arraystretch}{1.2}
    \small
    \centering
    \begin{tabular}{ll} 
    \toprule
    \textbf{Category} & \textbf{References} \\ 
    %  [0.5ex] 
    \midrule
    Airline Routes & \cite{lordan2014study,mattsson2015vulnerability,lordan2016robustness,du2016analysis,lordan2015robustness,zhou2019efficiency} \\
    Electrical Grid & \cite{pagani2013power,cuadra2015critical,bernstein2014power,holmgren2006using,kinney2005modeling,nguyen2013detecting,schneider2011mitigation} \\
    Road Networks & \cite{berdica2002introduction,derrible2010complexity,duan2014robustness,mattsson2015vulnerability,berche2009resilience} \\
    Social Networks & \cite{nguyen2021modularity,boldi2011robustness,moore2021inclusivity,colladon2017robustness,boldi2013robustness} \\
    Telecommunication & \cite{rueda2017robustness,bilal2013characterization,manzano2013connectivity,de2016reliability,schneider2011mitigation,cohen2011resilience} \\
    Water Distribution & \cite{yazdani2012applying,nardo2018applications,di2018applications,yazdani2010robustness,yazdani2011complex} \\
    \bottomrule
    \end{tabular}
    \caption{We organize a few key robustness works across six practical application domains, allowing the reader to quickly access relevant research.
    }
    \label{tab:measure-applications}
\end{table}

\subsection{Selecting a Robustness Measure}\label{sec:selecting-measures}
In this section we explore which robustness metrics may be well suited to particular types of network data through an analysis of two high impact use case scenarios---(i) transportation networks and (2) water distribution networks. 
We begin by exploring the domain specific robustness problems in each use case scenario, and identify important network properties.
Studying these network properties in conjunction with the current methods of robustness analysis for the domain, we highlight alternative robustness measures (based on Table~\ref{table:metric_comparison}) that have potential to improve performance for the task at hand. 
Through this process we take a first step in addressing how to select a robustness measure by providing a template for future domain specific network robustness analysis.

\bigskip
\noindent\textbf{Scenario 1: Transportation Networks.}
Societal dependence on transportation networks (e.g., roadway) is enormous, and only increasing in number and complexity.
Unfortunately, these networks are often purposefully designed to operate with minimal redundancy and high capacity in order to minimize costs.
As a result, they are extremely sensitive to failures and disruptions~\cite{mattsson2015vulnerability}.
As such, understanding and studying the vulnerability and robustness of these networks is critical. 

While there is currently no definitive definition of transport system vulnerability~\cite{mattsson2015vulnerability}, a well received and representative one is suggested by Berdica~\cite{berdica2002introduction}: 
``\emph{Vulnerability in the road transportation system is a susceptibility to incidents that can result in considerable reductions in road network serviceability}''.
In order to understand the nature and extent of the vulnerability posed to transportation networks, graph theoretic measures have evolved as a natural tool of representation.
In the study of transportation systems it has been suggested robustness measures take into account factors like:

\begin{enumerate}[label=(\roman*), labelindent=3mm, align=left, topsep=4pt, leftmargin=*, itemsep=4pt, widest=iii]
    \item \textit{average distance} between different stops~\cite{mattsson2015vulnerability}
    \item \textit{alternative pathways} of transport~\cite{derrible2010complexity}
    \item \textit{bottlenecks} inside and between communities~\cite{duan2014robustness}
\end{enumerate}

Currently, highly interpretable network analysis tools such as average distance, betweenness centrality, clustering coefficient and largest connected component are proposed as measures of system robustness~\cite{duan2014robustness,mattsson2015vulnerability}.
From a decision making perspective, these measures are highly interpretable helping inform decision making policy.

\medskip
\noindent\textbf{Potential Alternatives.}
Traditional robustness analysis applied to transportation networks could benefit from the use of spectral based techniques, which are scalable to large networks and could potentially provide better performance.
In this domain we identified alternative pathways (i.e. redundancy) and bottlenecks as important criteria in robustness measure selection.
Therefore the \textit{spectral gap} which accounts for bottlenecks, and \textit{effective resistance} which takes into account alternative pathways and their length, may be two important measures for robustness analysis.

\bigskip
\noindent\textbf{Scenario 2: Water Distribution Networks.}
Cities and municipalities depend on a mixture of complex and interconnected infrastructure to provide a reliable and safe source of water to consumers.
A serious concern for the water utilities providing this service is the vulnerability of water distribution networks (WDNs) to common failures (e.g., aging pipes) and targeted attacks (e.g., disrupted service)~\cite{yazdani2010robustness,di2018applications}.
Due to WDNs natural graph representation, graph robustness measures have become a critical tool in the analysis of network vulnerability.
In order to address the concern of vulnerability, multiple criteria have been proposed to evaluate a WDNs robustness:

\begin{enumerate}[label=(\roman*), labelindent=3mm, align=left, topsep=4pt, leftmargin=*, itemsep=4pt, widest=iii]
    \item \textit{alternative pathways} of supply~\cite{yazdani2011complex}
    \item \textit{bottlenecks} or articulation points~\cite{nardo2018applications}
    \item \textit{connectedness} of the network~\cite{nardo2018applications}
    \item \textit{loop redundancy}~\cite{yazdani2010robustness}
\end{enumerate}

Interestingly, compared to transportation networks, WDNs currently use a mixture of graph and spectral approaches to account for these desired properties.
Common analysis tools include average distance, diameter and clustering coefficient~\cite{yazdani2010robustness}, which could be used to identify graph connectedness and loop redundancy.
In addition spectral approaches such as spectral gap and algebraic connectivity have been proposed for use in WDN vulnerability analysis.
These techniques could be used to identify bottlenecks and measure the strength of connectivity between subregions~\cite{nardo2018applications}.

\medskip
\noindent\textbf{Potential Alternatives.}
Bottleneck and loop redundancy are important criteria in robustness measure selection.
Measures such as the spectral gap and algebraic currently can account for bottlenecks; while tools like average distance, diameter and clustering coefficient provide a measure of alternative pathways and loop redundancy. 
However in this case, \textit{natural connectivity} which is intrinsically tied to loop capacity and alternative pathways presents a compelling robustness measure.
In addition, \textit{effective resistance} which takes into account alternative pathways and their length could be a strong alternative measure.

\section{Failures and Targeted Attacks}\label{section:attack}
To understand the underlying mechanisms of network failure and attack, we need to examine the graph properties facilitating these issues.
In order to do this, we begin with a brief overview of four classical graph models in Section~\ref{subsec:graph-models}.
This background knowledge on graph models assists in the analysis of network failure and attack in Section~\ref{subsec:failures} and Section~\ref{subsec:targeted-attacks}, respectively.

\subsection{Graph Models}\label{subsec:graph-models}
We discuss a few key graph generators to provide the reader background on commonly studied models in attack analysis.
By understanding the robustness of graph models---which are representative of many real-world datasets---we can better understand new types of graph data that follow the same distributions.
For example, many real-world graphs are scale-free in nature~\cite{ebel2002scale,faloutsos2011power,jeong2000large,mislove2007measurement,albert2005scale}, meaning that once we determine a network is scale-free~\cite{clauset2009power}, we can leverage knowledge of scale-free models to better understand the robustness characteristics of the new graph.
There are also multiple advantages of modeling real-world graphs using a generator~\cite{akoglu2009rtg}, including---%
(1) \textit{sampling}, where we can create small graphs to quickly run a suite of robustness measures and attack/defense simulations; 
(2) \textit{extrapolation}, to evaluate the robustness characteristics of a graph in the future;
and (3) simulation studies, so users can quickly study robustness across various graph topological structures 

\medskip

\noindent\textbf{Erd\"{o}s-R\'{e}nyi (ER) Model}~\cite{erdos1960evolution}
The ER model generates random networks with no particular structural bias, where each graph starts with $n$ vertices and no edges. 
For each pair of nodes, an edge is added to the graph with probability $p$.
This leads to the Poisson-type degree distribution where the probability of a vertex having degree $k$ is $p_k = p(n-1)$~\cite{chakrabarti2006graph}.
In addition, the ER model has a logarithmically increasing average distance and a clustering coefficient close to zero~\cite{holme2002attack}.
Potentially the most important property of Erd\"{o}s-R\'{e}nyi graphs in relation to network robustness is the homogeneity of the degree and betweenness distributions.
This property of homogeneity implies that node importance and network load are evenly distributed among the nodes in the graph~\cite{xia2010cascading}. 
As we will see in the following sections, this is a particularly valuable property.

\medskip

\noindent\textbf{Watts-Strogatz (WS) Model}~\cite{watts1998collective}
The WS model generates graphs with high clustering coefficient and low diameter (small-world property).
The model starts starts by generating a ring lattice with $n$ vertices, where every node is connected to its $k$ nearest neighbors. 
Each edge is then visited once and rewired with probability $p$ to a vertex chosen uniformly at random (no duplicate edges or self-loops allowed).
This has the effect of creating ``shortcuts'' across the graph, creating the low diameter, small-world property.
As such, the WS model shows a heterogeneous betweenness distribution where a small number of nodes have very large betweenness while most nodes contain very little (not a power law distribution though).
However, the WS model does manage to maintain a homogeneous degree distribution due the small number of edges rewired~\cite{xia2010cascading}.

\medskip

\noindent\textbf{Barb\'{a}si-Albert (BA) Scale-Free Model}~\cite{barabasi1999emergence}
The BA model generates network topology according to two processes---\textit{growth} and \textit{preferential attachment}. 
Where prior network models kept the number of nodes fixed during the network formation process, the BA model starts with a small set of vertices and \textit{grows} the network by adding nodes and edges over time.
The reason why the BA model follows the sociological principle of the ``rich get richer'', where the probability of connecting to a node is proportional to the degree of that node, is due to \textit{preferential attachment} mechanism~\cite{chakrabarti2006graph}.
The BA model generates scale-free networks following a power law degree distribution given by $p_k \approx k^{-3}$ with average geodesic distance increasing logarithmically with the size of $n$, demonstrating the small-world property~\cite{holme2002attack}.
As such, the BA model shows both a heterogeneous degree and betweenness distribution, where a small number of nodes account for large proportion of the betweenness load and degree~\cite{xia2010cascading}.

\medskip

\noindent\textbf{Clustered Scale-Free (CSF) Model}~\cite{holme2002growing}
The CSF model extends the BA model by incorporating a high clustering coefficient through a single step after preferential attachment, called \textit{triad formation}.
The advantage of the CSF model compared to the BA model is that it generates graphs with high clustering coefficient while maintaining the scale-free and small average distance properties.
As such, the degree distribution of the CSF model is heterogeneous and roughly equivalent to the BA model~\cite{holme2002growing}.

\subsection{Isolated \& Cascading Failures}\label{subsec:failures}
These types of failures in a network often occur when a piece of equipment breaks down due to natural causes. 
In the study of graphs, this could correspond to the removal of either a node or an edge.
While random and cascading failures do occur, they are often less severe than targeted attacks.
In fact, \cite{xia2010cascading} shows that cascading failures are significantly less impactful than targeted attacks across a range of graph models (ER, WS and BA).
For both ER and BA graphs, the network damage from cascading failures is minimal, as shown in the analytical framework by \cite{cohen2011resilience}. 
While the WS model suffered significant damage from the cascading failure, it was still significantly less than the targeted attack.
As a whole, this indicates that isolated and cascading failures are less threatening than targeted attacks.
As such, we focus on targeted attacks.

\subsection{Targeted Attacks}\label{subsec:targeted-attacks}
There are two primary mechanisms an adversary can use to attack a network---\textit{removal of nodes} and \textit{removal of edges}~\footnote{We note the possibility of adding edges to disrupt a network, but are not aware of any research in this direction.}.
The goal of the attacker is to select nodes and edges considered important to the functionality of the network (e.g., critical power substations or electrical lines).
To accomplish this, an attacker typically relies on measures of node and edge centrality.
While any centrality measure can be used to generate a list of the top $k$ nodes and edges to remove, we focus on two traditional measures---degree and betweenness centrality.
Through these two measure, there are four common attack strategies: initial degree removal (ID), initial betweenness removal (IB), recalculated degree distribution removal (RD) and recalculated betweenness removal (RB)~\cite{holme2002attack}. 
Each of these attack strategies are discussed below.

\medskip
\noindent\textbf{Initial Degree Removal (ID)}
In this attack scenario each node $v$ in the network is ranked according to its degree $\delta_v$.
For a given budget $k$, an attacker removes vertices one by one starting with the highest degree nodes.
This has the effect of reducing the total number of edges in the network as fast as possible~\cite{holme2002attack}.
Since this attack only considers its' local neighborhood when making a decision, it is considered a \textit{local attack}.
The benefit of this locality is that the attack strategy is quick to compute; linear in the size of nodes in the graph.
The same strategy can be applied to the removal of edges from the network, where edge degree $\delta_e$ can be defined in several ways. Four common ones are shown below~\cite{holme2002attack}:

\begin{subequations}\label{eq:edge-degree}
\begin{align}
    \begin{split}
        \delta_e = \delta_u\cdot \delta_v
    \end{split}\\
    \begin{split}
        \delta_e = \delta_u + \delta_v
    \end{split}\\
    \begin{split}
        \delta_e = \min(\delta_u, \delta_v)
    \end{split}\\
    \begin{split}
        \delta_e = \max(\delta_u, \delta_v)
    \end{split}
\end{align}
\end{subequations}

\noindent In practice, Equation~\ref{eq:edge-degree} (a) has been shown to be the most effective method of defining edge degree in relation to attack strategy, since it correlates best with betweenness centrality~\cite{holme2002attack}.

\medskip
\noindent\textbf{Initial betweenness Removal (IB)}
This attack ranks each node $v\in V$ according to its betweenness centrality $b_v$ in Equation~\ref{eq:vertex-betweenness}.
The attacker then removes up to $k$ nodes, one at a time in descending order of importance.
This has the effect of destroying as many paths as possible~\cite{holme2002attack}.
Since this attack considers information from across the network, it is considered a \textit{global attack} strategy.
As a result, the IB attack strategy is significantly more computationally expensive than ID (see Section~\ref{subsec:edge-betweenness}).
The same attack strategy can be applied to the removal of edges using the definition of edge betweenness centrality:

\medskip
\noindent\textbf{Recalculated Degree Removal (RD)}
Recalculated degree removal follows the same steps as ID, except that it recalculates the degree distribution of nodes after removing each vertex.
This allows for the attacker to re-assess the target after each round of attack.
The same process holds for removing edges.

\medskip
\noindent\textbf{Recalculated betweenness Removal (RB)}
Recalculated betweenness removal follows the same steps as IB, except that it recalculates node centrality of each node is removed.
The same process holds for removing edges.

\subsubsection{Comparison of Targeted Attacks}\label{subsec:comparing-targeted-attacks}
The efficacy of each attack outlined in Section~\ref{subsec:targeted-attacks} is discussed in relation to two robustness measures: size of the largest connected component ($L$) and average inverse distance $\bar{d}^{-1}$, on four classic graph models---(1) Erd\"{o}s-R\'{e}nyi (ER), (2) Watts-Strogatz (WS), (3) Barab\'{a}si-Albert (BA) scale-free and (4) Clustered Scale-Free Model (CSF).
We summarize the effectiveness of each attack in Table~\ref{tab:attack-comparison} and provide a detailed review of attack efficacy on each type of graph below. 

\medskip

\noindent\textbf{Attacking Erd\"{o}s-R\'{e}nyi Graphs.}
The most effective node attack strategy on ER graphs is RD.
However, it takes the removal of $\sim$30\% of the most central nodes in the graph in order to reduce the $\bar{d}^{-1}$ by 50\%; and the removal of $\sim$40\% of the central nodes to decrease $L$ by 50\%~\cite{holme2002attack}.
This is a large fraction of the nodes to be removed, indicating that ER graph are highly robust to targeted attacks.
This robustness of Erd\"{o}s-R\'{e}nyi graphs stem from the homogeneous degree and betweenness distributions.
Since these distributions evenly spread the load and importance among each of the nodes. 
Therefore if a few nodes are attacked, no serious network damage occurs~\cite{xia2010cascading}.

The ER model is also robust to edge based attacks. 
The most effective attack strategy is again RB, where approximately 40\% of the edges have to be removed in order to drop the average inverse distance by 50\%.
To reach a similar performance drop in the largest connected component, approximately 50\% of the edges needed to be removed by the most effective edge attack strategy RD~\cite{holme2002attack}. 

\medskip

\noindent\textbf{Attacking Watts-Strogatz Graphs.}
The attack behavior on WS graphs is significantly different from ER graphs.
For just a small number of removed vertices, the RB procedure completely breaks down the structure of the graph. 
By removing just $\sim$2\% of the most central nodes, $\bar{d}^{-1}$ is reduced by 45\%.
After removing roughly 10\% of the top nodes, the network structure completely breaks down resulting in a 95\% decrease in $\bar{d}^{-1}$ and an 85\% decrease in $L$~\cite{holme2002attack}.
However, results from \cite{holme2002attack} show that degree based strategies are largely ineffective in attacking WS graphs. 
This largely stem from the fact that WS networks have a heterogeneous betweenness distribution and a homogeneous degree distribution.
Since the RB attack targets central nodes carrying a majority of the load, the graph rapidly disintegrates.
On the other hand, targeting nodes based on degree distribution is not as effective since most nodes have roughly the same degree~\cite{xia2010cascading}.

Evaluating the robustness of WS graphs under edge attacks results in similar performance degradation compared to node based ones.
Again, the RB attack effectively deconstructs the graph after the removal of only 20\% of the edges, reducing $L$ to $<5\%$ of its original value. 
In addition, the average inverse distance drops to roughly 10\% of the original value at the same mark.
The effectiveness of the RB strategy can be tied to the identification of important edges, which in the WS model are the rewired edges linking the distant parts of the ring~\cite{holme2002attack}.
It is interesting to note that the RD strategy is significantly less effective than ID when removing edges---an interesting discussion on this surprising result can be found in \cite{holme2002attack}. 

\medskip

\noindent\textbf{Attacking Barb\'{a}si-Albert Scale-Free Graphs.}
The BA graph is significantly more robust to node based attacks than the WS model, even though the BA graph has a heterogeneous degree and betweenness distribution. 
While this may seem odd at first, it likely stems from the fact that the WS model is based on a ring structure with local connections, where the removal of even a single node significantly affects the neighborhood~\cite{xia2010cascading}.
While BA graphs are overall quite robust, compared to the ER model, the BA graph is more vulnerable to both RB and RD attacks due to the heterogeneous nature of the degree and betweenness distribution.

Edge based attacks on the BA model perform poorly compared to their node based counterparts.
It takes a removal of 40\% of the edges in order to drop the performance of $\bar{d}^{-1}$ by 50\%. In comparison, it would have only the removal of $\sim$12\% of the nodes to reach the same level of performance drop.
Similar results are found when comparing the largest connected component performance across node and edge attacks~\cite{holme2002attack}.
One potential reason for this significant difference could arise from the combination of the homogeneous nature of the betweenness distribution and the hub-like structure of important vertices.

\medskip

\noindent\textbf{Attacking Clustered Scale-Free Graphs.}
The CSF model is an extension of BA with the addition of a triad step, allowing the model to generate graphs with a higher clustering coefficient. 
However, it turns out that this clustering step introduces significant vulnerability into the network.
This vulnerability is likely caused by the targeting of important vertices with many triads attached to it~\cite{holme2002attack}.
To put it in comparison, removing just 10\% of the central nodes from the CSF graph results in both $L$ and $\bar{d}^{-1}$ dropping by 90\%.
In comparison, the BA graph which is widely known to be vulnerable to targeted attacks, only drops roughly 50\% in $L$ and 10\% in $\bar{d}^{-1}$ with the same number of nodes removed~\cite{holme2002attack}.

Similar to the BA graph, CSF graphs seem to be more resilient to edge based attacks than node based ones~\cite{holme2002attack}.
However, it is still significantly more vulnerable to RD based edge attacks than the BA model.

\begin{table}[t]
    \setlength{\tabcolsep}{10pt}
    \renewcommand{\arraystretch}{1.2}
    \small
    \centering
    \begin{tabular}{llrrrrr} 
    \toprule
    & \textbf{Attack} & \textbf{ER} & \textbf{WS}  & \textbf{BA}  & \textbf{CSF} \\ 
    %  [0.5ex] 
    \midrule
    \multirow{4}{*}{\textbf{Node}}
    & ID & \xmark & \cmark & \cmark & \cmark \\
    & IB & \xmark & \cmark & \cmark & \cmark \\
    & RD & \xmark & \cmark & \cmark & \cmark \\
    & RB & \xmark & \cmark & \cmark & \cmark \\
    \midrule
    \multirow{4}{*}{\textbf{Edge}}
    & ID & \xmark & \cmark & \xmark & \xmark \\
    & IB & \xmark & \cmark & \xmark & \xmark \\
    & RD & \xmark & \xmark & \xmark & \xmark \\
    & RB & \xmark & \cmark & \xmark & \cmark \\
    \bottomrule
    \end{tabular}
    \caption{Comparing the effectiveness of node and edge attack strategies against classic graph models~\cite{holme2002attack}.
    A check mark (\cmark) and cross (\xmark) indicates if an attack is a generally effective or ineffective attack strategy, respectively.
    See Section~\ref{subsec:comparing-targeted-attacks} for a more fine grained comparison.
    }
    \label{tab:attack-comparison}
\end{table}

\subsection{Comparison to Other Targeted Attacks}
While we have studied two types of centrality based attacks on networks---node centrality and betweenness centrality---there exists a number of alternative options~\cite{baig2015correlation}.
Some of these alternatives include: PageRank~\cite{page1999pagerank}, closeness centrality~\cite{opsahl2010node}, eigenvector centrality and Katz centrality~\cite{katz1953new}.
However, it has been shown that many of these centrality scores produce highly correlated results when conducting targeted attacks on networks~\cite{baig2015correlation}.
The three most common groupings of centrality measures according to similarity are as follows: (PageRank, betweenness centrality), (Katz centrality, eigenvector centrality) and (closeness centrality, degree centrality).
As such, attack related studies may want to consider evaluating attacks from distinct groups in order to avoid similar attack patterns.

\section{Network Defense}\label{section:defense}
To understand the best mechanism of defense for a particular network, it is important to understand the properties of the graph and the type of attack or failure we are expecting to protect against.
In Section~\ref{subsec:defense-heuristic} we overview measure independent heuristics for improving network defense, grouping each technique into one of three categories depending on whether it improves network robustness through (i) edge addition, (ii) edge rewiring, or (iii) identifies important nodes and edges in a network to monitor suspicious activity. 
Then in Section~\ref{subsec:defense-opt} we analyze optimization based techniques for network defense according to the same categorization process as above.
Finally, in Section~\ref{subsec:defense-selection} we discuss when to apply different defense techniques.

\subsection{Measure Independent Heuristics}\label{subsec:defense-heuristic}
In this section we evaluate network defenses that are heuristic in nature. This means that the technique modifies the graph structure independent of a robustness measure~\cite{chan2016optimizing}.

\subsubsection{Edge Addition}\label{subsubsec:edge-addition}
Edge additions typically incur additional network costs then edge rewiring, however, they almost always provide better results.
In~\cite{beygelzimer2005improving}, they show that edge addition outperforms all proposed edge rewiring schemes on two measures of network robustness $L$ and $\bar{d}^{-1}$. 
In addition, they find that not all edge addition techniques are equal, and that preferential edge addition outperforms random edge addition:

\begin{enumerate}[label=(\arabic*), labelindent=3mm, align=left, topsep=4pt, leftmargin=*, itemsep=4pt]
    \item \textit{Random addition}: add an edge connecting two random nodes
    
    \item \textit{Preferential addition}: add an edge by connecting two nodes having the lowest degrees
\end{enumerate}

\subsubsection{Edge Rewiring}\label{subsubsec:edge-rewiring}
Edge rewiring is a popular technique to improve network robustness since it generally has a lower cost associated with it compared to adding edges.
As such, we study the 4 methods of edge rewiring proposed in~\cite{beygelzimer2005improving}:

\begin{enumerate}[label=(\arabic*), labelindent=3mm, align=left, topsep=4pt, leftmargin=*, itemsep=4pt]
    \item \textit{Random edge rewiring}: remove a random edge, then add an edge as in (1)
    
    \item \textit{Random neighbor rewiring}: randomly select a node, and then a random neighbor of that node, then remove the corresponding edge. Next add an edge as in (1)
    
    \item \textit{Preferential rewiring}: disconnect a random edge from the highest-degree node and reconnect that edge to a random node
    
    \item \textit{Preferential random edge rewiring}: choose a random edge, disconnect it from the higher degree node, then connect that edge to a random node
\end{enumerate}

In order to evaluate the effectiveness of each proposed rewiring scheme, \cite{beygelzimer2005improving} proposes to measure network robustness through the average inverse distance $\bar{d}^{-1}$ and largest connected component $L$ on three increasing levels of node attack.
With respect to robustness measure $L$, the conclusion that was drawn is that edge rewiring is most effective in the following order: $5 > (3 \text{ and } 6) > 4$. Each number corresponds to the list above.
On the other hand when using $\bar{d}^{-1}$ is used as the metric, preferential rewiring (6) is found to increase $\bar{d}^{-1}$ the most for small amounts of rewiring, while random edge (3) and random neighbor (4) techniques give the largest $\bar{d}^{-1}$ improvement when large numbers of edges were rewired~\cite{beygelzimer2005improving}.

\subsubsection{Node Monitoring}\label{subsubsec:node-monitoring}
Many of the centrality based techniques that are used to attack a network can be used to defend it.
In fact, if you know that an adversary is using one of these approaches to attack your network, you can monitor the same nodes in advance.
To this end, a range of centrality measures can be used for node monitoring, including: degree centrality, betweenness centrality, PageRank, closeness centrality, eigenvector centrality, Katz centrality, etc.
While picking the appropriate measure to monitor your network might seem overwhelming, it has been shown that a high degree of correlation exists between many of these measures.
The three most common groupings of centrality measures are: (PageRank, betweeness centrality), (Katz centrality, eigenvector centrality) and (closeness, degree)~\cite{baig2015correlation}.
This can help reduce the burden of selecting centrality measures for defense by eliminating redundant and highly correlated measures.

\subsection{Optimization Based Techniques}\label{subsec:defense-opt}
The focus of optimization based methods is to modify the underlying graph structure through the manipulation of a targeted robustness measure~\cite{chan2016optimizing}.

\subsubsection{Edge Rewiring}
For this section, we focus our attention to the optimization based method for edge rewiring proposed in~\cite{chan2016optimizing}.
In this work, they propose an algorithm \textsc{EdgeRewire} that maximizes a particular spectral robustness measure according to a given budget. 
In addition, they only allow modifications based on degree-preserving edge rewirings in order to ensure that load on nodes remains unchanged.
In total, they allow \textsc{EdgeRewire} to operate on six spectral measures, including: spectral radius, spectral gap, natural connectivity, algebraic connectivity, effective resistance and number of spanning trees.

Comparing \textsc{EdgeRewire} to a series of heuristic based edge rewiring approaches on 14 datasets, they find that the proposed method significantly outperforms heuristic based approaches when measuring for the optimized spectral parameter.

\subsubsection{Node Monitoring}
We focus our study of optimization based node monitoring techniques to the work conducted by Tong et. al.~\cite{tong2010vulnerability}.
In this work, they propose a three step process for network defense against virus propagation---(i) evaluation of a graphs' vulnerability to virus propagation via the spectral radius; 
(ii) design of the `Shield-value' measure to quantify the importance of a set of nodes in protecting the graph; 
and (iii) development of a quick algorithm utilizing the Shield-value measure to determine the $k$ best nodes to protect the graph.

The spectral radius $\lambda_1$ was chosen as a natural measure of graph robustness to virus propagation due to its close link to the epidemiological threshold.
As such, in order to minimize the spread of a virus on a network the goal is to minimize $\lambda_1$ by selecting the best set $S$ of $k$ nodes to remove from the graph (i.e., maximize eigendrop). 
In order to evaluate the goodness of a node set $S$ for removal, \cite{tong2010vulnerability} proposes the Shield-value measure:

\begin{equation}
    Sv(S) = \sum_{i\in S}2\lambda_1 \bm{u}_1(i)^2 - \sum_{i,j\in S} A(i,j)\bm{u}(i)\bm{u}(j)
\end{equation}

The intuition behind this equation is that we want to select nodes for removal/monitoring that have high eigenvector centrality (first term) while penalizing nodes for being close together (second term).
In order to select this set of nodes $S$, they develop the NetShield algorithm. We refer the reader to~\cite{tong2010vulnerability} for technical details.
To evaluate the efficacy of the node monitoring approach, it is compared across a multitude of heuristic based measures, including--- PageRank, degree centrality, etc.---finding that the NetShield approach to node monitoring outperforms traditional centrality based approaches in mitigating the spread of viruses on a network.

\subsection{Selecting a Defense Method}\label{subsec:defense-selection}
The most devastating attacks often correspond to targeted attacks, rather than isolated or cascading failures~\cite{xia2010cascading}.
As a result, it is common for defense mechanisms to prioritize the protection of networks from targeted attacks, unless prior information indicates otherwise.
Since an attacker likely targets nodes or edges based on a measure of centrality, we have information \textit{a priori} on our adversary.
As the defender, we often have knowledge of the graph topology and underlying degree distribution.
This allows us to better quantify our robustness and identify potential points of attack
For example, we can measure the number of bottlenecks present in our network; along with the availability of alternative pathways.
Perhaps most importantly, we can determine the degree and betweenness distribution of the graph topology, allowing us to make additional assumptions about the best defense strategy.

Since many real world networks assume a heterogeneous degree distribution, where a few nodes contain many links and many nodes contain only a few links; we use it as an example network to defend.
In this scenario we can: (1) monitor critical nodes according to different measures of centrality (e.g., nodes with many links); (2) rewire the graph in an attempt to increase robustness; or (3) add edges to the network in order to increase robustness.
In practice, these decisions often depend on the specific application domain and the cost of the associated action.
However, the most cost effective measure is arguably node monitoring, which if implemented carefully, can prevent targeted attacks from ever occurring. 

\section{Research Directions \& Open Problems}\label{sec:future}
We present research directions and open problems distilled from the surveyed works.
Four promising directions include: 
(1) an axiomatic study of desired properties in robustness measures, helping guide the selection and development of new measures;
(2) interpretability of robustness measures to assist users in understanding the impact of measure scores;
and (3) applying the study of network robustness to additional high-impact domains such as physical security and cybersecurity.
% and (4) bridging the study of graph vulnerability and robustness with recent developments in adversarial machine learning on graph structured data.   

\subsection{Guidelines for Selecting \& Developing Measures}
Comparing robustness measures in a quantitative manner is still an open challenge.
While many works have qualitatively remarked on why certain robustness measures are better suited for certain tasks, there has been no formal study outlining desirable characteristics that a robustness measure should contain.
By formalizing these desirable properties into a set axioms, future and existing robustness measures could be compared in an independent and quantitative manner, something that is not currently available.
We identified $6$ desirable robustness properties across the literature that could form the basis for an axiomatic analysis of robustness measures~\cite{ellens2013graph,chan2016optimizing,chan2014make}.
Below, we provide the intuition for each axiom, however, each axiom needs to be formalized and (dis)proven for each robustness measure.

\medskip
\noindent\textbf{1. Strictly Monotonic.} When an edge is added to a graph the network connectivity is intrinsically enhanced.
A robustness measure should account for this increased connectivity by strictly increasing (or decreasing) for each edge added to the graph.

\medskip
\noindent\textbf{2. Redundancy.}
A critical ability of any robustness measure is to measure redundancy present in the network.
This means that if multiple paths between two nodes exist, the proposed measure should be able to account for both the number of paths and their quality (where smaller paths are better).

\medskip
\noindent\textbf{3. Disconnected.}
Many real-world graphs contain disconnected components; therefore a measure should be able to evaluate a graphs' robustness independent of the number of disconnected components.

\medskip
\noindent\textbf{4. Stable.}
A robustness measure should change in proportion to the perturbation of the graph structure. 
For example, if a single edge is added to a graph, we expect that the measure has a proportionally small response.

\medskip
\noindent\textbf{5. Consistent.}
Given two graphs with same underlying structure, we would expect them to have similar robustness independent of their size.

\medskip
\noindent\textbf{6. Scalable.}
Large graphs containing millions (or sometimes billions) of nodes and edges are common.
A robustness measure should be scalable to large graphs, where we define scalable as an algorithm subquadratic with respect to the number of nodes and edges.

\medskip
\noindent\textbf{6. Intuitive.}
Ideally, we want robustness measures to have identifiable connections to the underlying graph topology, and for these connections to be conveyable to non-experts in an understandable manner.

\subsection{Furthering Interpretability}
Ideally, robustness measure should have identifiable connections to the underlying graph topology to explain what the robustness score is indicative of.
Recent research has explored this in the more general domains of graph connectivity and ranking~\cite{kang2018aurora,wang2020graph,kang2018x,xie2020auditing}.
Combining visual representations, helpful interactions, and state-of-the-art attribution and feature visualization techniques together into rich user interfaces could lead to major breakthroughs in understanding graph vulnerability and robustness scores.

\subsection{Studying Robustness in New Domains}
The study of graph vulnerability and robustness is still nascent in the areas of physical security~\cite{byrne2005algebraic}, cybersecurity~\cite{freitas2020d2m} and interdependent and dynamic networks~\cite{chen2018realtime}.
For physical security, \cite{byrne2005algebraic} studies the vulnerability and robustness of physical sensor placement to maximize perimeter security while minimizing network latency.
They find that perimeter security systems frequently map to circular lattices which suffer from a trade-off between robustness and mean path length (i.e., latency).
Future work could analyze alternative perimeter system mappings that optimize for both criteria, while exploring alternative definitions of robustness in physical security.
With respect to cybersecurity, \cite{freitas2020d2m} attempts to calculate the vulnerability and robustness of enterprise networks by modeling lateral attack movement between computers.
However, their unique probabilistic robustness measure is dependent on cyber domain knowledge and Monte Carlo simulations.
Future cybersecurity robustness work could explore the development of robustness measures that are simulation independent, reducing computational costs and the need for explicit domain knowledge.
In addition, many real-world networks are dynamic, spatial, and contain interdependent sub-networks.
While initial work has been done to introduce an analytical framework for interdependent~\cite{buldyrev2010catastrophic,chen2018realtime} spatial networks~\cite{danziger2016effect,barthelemy2011spatial}---additional work needs to be done to (i) study dynamic graphs, (ii) comprehensively evaluate various attack and defense scenarios, and (iii) develop unique robustness measures that can better account for the nature of interdependent, spatial, and dynamic networks.

% \subsection{Bridging Graph Robustness \& Adversarial ML}
% From the machine learning perspective, a majority of current graph robustness research focuses on manipulating graph classifiers or embedding mechanisms into mispredicting the label of a graph~\cite{dai2018adversarial}, or the label of each node in the graph~\cite{zugner2018adversarial}.
% So far, adversarial machine learning research has yet to deeply delve into the more richly defined network robustness objective centered around a networks' ability to continue functioning when damaged or attacked.
% However, we believe that there are multiple high-impact connections to explore, including: 
% (1) how does a graph's spectral robustness (e.g., spectral gap) correlate to the vulnerability or robustness of downstream tasks such as node and graph classifiers being attacked;
% and (2) does optimizing a graph's spectral robustness (e.g., adding or rewiring edges) affect an attackers ability to perturb downstream node and graph classification models.
% By answering these questions, we can uncover new mechanisms to attack and defend networks while gaining insight into fundamental connections between two important and growing fields.
\section{Conclusion}\label{sec:conclusion}

In this survey, we distill answers to key questions that are currently scattered across multiple scientific fields and numerous papers.
In particular, we provide researchers and practitioners crucial access to network robustness information information by---%  
(1) summarizing and comparing 17 recent and classical graph robustness measures; 
(2) exploring which robustness measures are most applicable to different categories of network data (e.g., social, infrastructure);
(3) reviewing common network attack strategies, and summarizing which attacks are most effective across different network topologies; and
(4) extensive discussion on selecting defense techniques to mitigate attacks across a variety of networks.
We concluded by highlighting current research directions and open problems.
This survey will serve as a guide to researchers and practitioners in navigating the expansive field of network robustness, while summarizing answers to key questions.
\section{Acknowledgements}
This work was in part supported by the NSF grant IIS-1563816, GRFP (DGE-1650044), a Google grant, a Raytheon fellowship, and an IBM fellowship.

\bibliographystyle{IEEEtran}
\bibliography{main.bib}

\begin{thebibliography}{100}
\providecommand{\url}[1]{#1}
\csname url@samestyle\endcsname
\providecommand{\newblock}{\relax}
\providecommand{\bibinfo}[2]{#2}
\providecommand{\BIBentrySTDinterwordspacing}{\spaceskip=0pt\relax}
\providecommand{\BIBentryALTinterwordstretchfactor}{4}
\providecommand{\BIBentryALTinterwordspacing}{\spaceskip=\fontdimen2\font plus
\BIBentryALTinterwordstretchfactor\fontdimen3\font minus
  \fontdimen4\font\relax}
\providecommand{\BIBforeignlanguage}[2]{{%
\expandafter\ifx\csname l@#1\endcsname\relax
\typeout{** WARNING: IEEEtran.bst: No hyphenation pattern has been}%
\typeout{** loaded for the language `#1'. Using the pattern for}%
\typeout{** the default language instead.}%
\else
\language=\csname l@#1\endcsname
\fi
#2}}
\providecommand{\BIBdecl}{\relax}
\BIBdecl

\bibitem{chvatal1973tough}
V.~Chv{\'a}tal, ``Tough graphs and hamiltonian circuits,'' \emph{Discrete
  Mathematics}, vol.~5, no.~3, pp. 215--228, 1973.

\bibitem{klein1993resistance}
D.~J. Klein and M.~Randi{\'c}, ``Resistance distance,'' \emph{Journal of
  mathematical chemistry}, vol.~12, no.~1, pp. 81--95, 1993.

\bibitem{beygelzimer2005improving}
A.~Beygelzimer, G.~Grinstein, R.~Linsker, and I.~Rish, ``Improving network
  robustness by edge modification,'' \emph{Physica A: Statistical Mechanics and
  its Applications}, vol. 357, no. 3-4, pp. 593--612, 2005.

\bibitem{tong2010vulnerability}
H.~Tong, B.~A. Prakash, C.~Tsourakakis, T.~Eliassi-Rad, C.~Faloutsos, and D.~H.
  Chau, ``On the vulnerability of large graphs,'' in \emph{2010 IEEE
  International Conference on Data Mining}.\hskip 1em plus 0.5em minus
  0.4em\relax IEEE, 2010, pp. 1091--1096.

\bibitem{krishnamoorthy1987fault}
M.~S. Krishnamoorthy and B.~Krishnamurthy, ``Fault diameter of interconnection
  networks,'' \emph{Computers \& Mathematics with Applications}, vol.~13, no.
  5-6, pp. 577--582, 1987.

\bibitem{ellens2013graph}
W.~Ellens and R.~E. Kooij, ``Graph measures and network robustness,''
  \emph{arXiv preprint arXiv:1311.5064}, 2013.

\bibitem{chan2016optimizing}
H.~Chan and L.~Akoglu, ``Optimizing network robustness by edge rewiring: a
  general framework,'' \emph{Data Mining and Knowledge Discovery}, vol.~30,
  no.~5, pp. 1395--1425, 2016.

\bibitem{yazdani2012applying}
A.~Yazdani and P.~Jeffrey, ``Applying network theory to quantify the redundancy
  and structural robustness of water distribution systems,'' \emph{Journal of
  Water Resources Planning and Management}, vol. 138, no.~2, pp. 153--161,
  2012.

\bibitem{lordan2014study}
O.~Lordan, J.~M. Sallan, and P.~Simo, ``Study of the topology and robustness of
  airline route networks from the complex network approach: a survey and
  research agenda,'' \emph{Journal of Transport Geography}, vol.~37, pp.
  112--120, 2014.

\bibitem{pagani2013power}
G.~A. Pagani and M.~Aiello, ``The power grid as a complex network: a survey,''
  \emph{Physica A: Statistical Mechanics and its Applications}, vol. 392,
  no.~11, pp. 2688--2700, 2013.

\bibitem{cuadra2015critical}
L.~Cuadra, S.~Salcedo-Sanz, J.~Del~Ser, S.~Jim{\'e}nez-Fern{\'a}ndez, and Z.~W.
  Geem, ``A critical review of robustness in power grids using complex networks
  concepts,'' \emph{Energies}, vol.~8, no.~9, pp. 9211--9265, 2015.

\bibitem{albert2000error}
R.~Albert, H.~Jeong, and A.-L. Barab{\'a}si, ``Error and attack tolerance of
  complex networks,'' \emph{nature}, vol. 406, no. 6794, pp. 378--382, 2000.

\bibitem{alenazi2015evaluation}
M.~J. Alenazi and J.~P. Sterbenz, ``Evaluation and comparison of several graph
  robustness metrics to improve network resilience,'' in \emph{2015 7th
  International Workshop on Reliable Networks Design and Modeling
  (RNDM)}.\hskip 1em plus 0.5em minus 0.4em\relax IEEE, 2015, pp. 7--13.

\bibitem{alenazi2015comprehensive}
------, ``Comprehensive comparison and accuracy of graph metrics in predicting
  network resilience,'' in \emph{2015 11th International Conference on the
  Design of Reliable Communication Networks (DRCN)}.\hskip 1em plus 0.5em minus
  0.4em\relax IEEE, 2015, pp. 157--164.

\bibitem{baig2015correlation}
M.~B. Baig and L.~Akoglu, ``Correlation of node importance measures: An
  empirical study through graph robustness,'' in \emph{Proceedings of the 24th
  International Conference on World Wide Web}, 2015, pp. 275--281.

\bibitem{baras2009efficient}
J.~S. Baras and P.~Hovareshti, ``Efficient and robust communication topologies
  for distributed decision making in networked systems,'' in \emph{Proceedings
  of the 48h IEEE Conference on Decision and Control (CDC) held jointly with
  2009 28th Chinese Control Conference}.\hskip 1em plus 0.5em minus 0.4em\relax
  IEEE, 2009, pp. 3751--3756.

\bibitem{berdica2002introduction}
K.~Berdica, ``An introduction to road vulnerability: what has been done, is
  done and should be done,'' \emph{Transport policy}, vol.~9, no.~2, pp.
  117--127, 2002.

\bibitem{bernstein2014power}
A.~Bernstein, D.~Bienstock, D.~Hay, M.~Uzunoglu, and G.~Zussman, ``Power grid
  vulnerability to geographically correlated failures—analysis and control
  implications,'' in \emph{IEEE INFOCOM 2014-IEEE Conference on Computer
  Communications}.\hskip 1em plus 0.5em minus 0.4em\relax IEEE, 2014, pp.
  2634--2642.

\bibitem{bigdeli2009comparison}
A.~Bigdeli, A.~Tizghadam, and A.~Leon-Garcia, ``Comparison of network
  criticality, algebraic connectivity, and other graph metrics,'' in
  \emph{Proceedings of the 1st Annual Workshop on Simplifying Complex Network
  for Practitioners}, 2009, pp. 1--6.

\bibitem{bishop2011link}
A.~N. Bishop and I.~Shames, ``Link operations for slowing the spread of disease
  in complex networks,'' \emph{EPL (Europhysics Letters)}, vol.~95, no.~1, p.
  18005, 2011.

\bibitem{boccaletti2006complex}
S.~Boccaletti, V.~Latora, Y.~Moreno, M.~Chavez, and D.-U. Hwang, ``Complex
  networks: Structure and dynamics,'' \emph{Physics reports}, vol. 424, no.
  4-5, pp. 175--308, 2006.

\bibitem{borgatti2006robustness}
S.~P. Borgatti, K.~M. Carley, and D.~Krackhardt, ``On the robustness of
  centrality measures under conditions of imperfect data,'' \emph{Social
  networks}, vol.~28, no.~2, pp. 124--136, 2006.

\bibitem{briesemeister2003epidemic}
L.~Briesemeister, P.~Lincoln, and P.~Porras, ``Epidemic profiles and defense of
  scale-free networks,'' in \emph{Proceedings of the 2003 ACM workshop on Rapid
  malcode}, 2003, pp. 67--75.

\bibitem{buldyrev2010catastrophic}
S.~V. Buldyrev, R.~Parshani, G.~Paul, H.~E. Stanley, and S.~Havlin,
  ``Catastrophic cascade of failures in interdependent networks,''
  \emph{Nature}, vol. 464, no. 7291, pp. 1025--1028, 2010.

\bibitem{byrne2005algebraic}
R.~Byrne, J.~Feddema, and C.~Abdallah, ``Algebraic connectivity and graph
  robustness,'' \emph{SANDIA Report}, vol. 87185, pp. 1--34, 2005.

\bibitem{caballero2008would}
J.~Caballero, T.~Kampouris, D.~Song, and J.~Wang, ``Would diversity really
  increase the robustness of the routing infrastructure against software
  defects?'' in \emph{NDSS}.\hskip 1em plus 0.5em minus 0.4em\relax Citeseer,
  2008.

\bibitem{callaway2000network}
D.~S. Callaway, M.~E. Newman, S.~H. Strogatz, and D.~J. Watts, ``Network
  robustness and fragility: Percolation on random graphs,'' \emph{Physical
  review letters}, vol.~85, no.~25, p. 5468, 2000.

\bibitem{chan2014make}
H.~Chan, L.~Akoglu, and H.~Tong, ``Make it or break it: Manipulating robustness
  in large networks,'' in \emph{Proceedings of the 2014 SIAM International
  Conference on Data Mining}.\hskip 1em plus 0.5em minus 0.4em\relax SIAM,
  2014, pp. 325--333.

\bibitem{chakrabarti2008epidemic}
D.~Chakrabarti, Y.~Wang, C.~Wang, J.~Leskovec, and C.~Faloutsos, ``Epidemic
  thresholds in real networks,'' \emph{ACM Transactions on Information and
  System Security (TISSEC)}, vol.~10, no.~4, pp. 1--26, 2008.

\bibitem{chen2015node}
C.~Chen, H.~Tong, B.~A. Prakash, C.~E. Tsourakakis, T.~Eliassi-Rad,
  C.~Faloutsos, and D.~H. Chau, ``Node immunization on large graphs: Theory and
  algorithms,'' \emph{IEEE Transactions on Knowledge and Data Engineering},
  vol.~28, no.~1, pp. 113--126, 2015.

\bibitem{chen2015connectivity}
C.~Chen, J.~He, N.~Bliss, and H.~Tong, ``On the connectivity of multi-layered
  networks: Models, measures and optimal control,'' in \emph{2015 IEEE
  International Conference on Data Mining}.\hskip 1em plus 0.5em minus
  0.4em\relax IEEE, 2015, pp. 715--720.

\bibitem{chen2016eigen}
C.~Chen, H.~Tong, B.~A. Prakash, T.~Eliassi-Rad, M.~Faloutsos, and
  C.~Faloutsos, ``Eigen-optimization on large graphs by edge manipulation,''
  \emph{ACM Transactions on Knowledge Discovery from Data (TKDD)}, vol.~10,
  no.~4, pp. 1--30, 2016.

\bibitem{crucitti2004model}
P.~Crucitti, V.~Latora, and M.~Marchiori, ``Model for cascading failures in
  complex networks,'' \emph{Physical Review E}, vol.~69, no.~4, p. 045104,
  2004.

\bibitem{dekker2005simulating}
A.~H. Dekker, ``Simulating network robustness for critical infrastructure
  networks,'' in \emph{ACM International Conference Proceeding Series}, vol.
  102.\hskip 1em plus 0.5em minus 0.4em\relax Citeseer, 2005, pp. 59--67.

\bibitem{derrible2010complexity}
S.~Derrible and C.~Kennedy, ``The complexity and robustness of metro
  networks,'' \emph{Physica A: Statistical Mechanics and its Applications},
  vol. 389, no.~17, pp. 3678--3691, 2010.

\bibitem{duan2014robustness}
Y.~Duan and F.~Lu, ``Robustness of city road networks at different
  granularities,'' \emph{Physica A: Statistical Mechanics and its
  Applications}, vol. 411, pp. 21--34, 2014.

\bibitem{ellens2011effective}
W.~Ellens, F.~Spieksma, P.~Van~Mieghem, A.~Jamakovic, and R.~Kooij, ``Effective
  graph resistance,'' \emph{Linear algebra and its applications}, vol. 435,
  no.~10, pp. 2491--2506, 2011.

\bibitem{estrada2006network}
E.~Estrada, ``Network robustness to targeted attacks. the interplay of
  expansibility and degree distribution,'' \emph{The European Physical Journal
  B-Condensed Matter and Complex Systems}, vol.~52, no.~4, pp. 563--574, 2006.

\bibitem{estrada2006spectral}
------, ``Spectral scaling and good expansion properties in complex networks,''
  \emph{EPL (Europhysics Letters)}, vol.~73, no.~4, p. 649, 2006.

\bibitem{freitas2020d2m}
S.~Freitas, A.~Wicker, D.~H. Chau, and J.~Neil, ``D2m: Dynamic defense and
  modeling of adversarial movement in networks,'' \emph{SDM}, 2020.

\bibitem{freitas2021evaluating}
S.~Freitas, D.~Yang, S.~Kumar, H.~Tong, and D.~H. Chau, ``Evaluating graph
  vulnerability and robustness using tiger,'' in \emph{Proceedings of the 30th
  ACM International Conference on Information \& Knowledge Management}, 2021,
  pp. 4495--4503.

\bibitem{gao2011robustness}
J.~Gao, S.~V. Buldyrev, S.~Havlin, and H.~E. Stanley, ``Robustness of a network
  of networks,'' \emph{Physical Review Letters}, vol. 107, no.~19, p. 195701,
  2011.

\bibitem{ghosh2008minimizing}
A.~Ghosh, S.~Boyd, and A.~Saberi, ``Minimizing effective resistance of a
  graph,'' \emph{SIAM review}, vol.~50, no.~1, pp. 37--66, 2008.

\bibitem{holme2002attack}
P.~Holme, B.~J. Kim, C.~N. Yoon, and S.~K. Han, ``Attack vulnerability of
  complex networks,'' \emph{Physical review E}, vol.~65, no.~5, p. 056109,
  2002.

\bibitem{holmgren2006using}
{\AA}.~J. Holmgren, ``Using graph models to analyze the vulnerability of
  electric power networks,'' \emph{Risk analysis}, vol.~26, no.~4, pp.
  955--969, 2006.

\bibitem{jamakovic2007relationship}
A.~Jamakovic and S.~Uhlig, ``On the relationship between the algebraic
  connectivity and graph's robustness to node and link failures,'' in
  \emph{2007 Next Generation Internet Networks}.\hskip 1em plus 0.5em minus
  0.4em\relax IEEE, 2007, pp. 96--102.

\bibitem{khalil2014scalable}
E.~B. Khalil, B.~Dilkina, and L.~Song, ``Scalable diffusion-aware optimization
  of network topology,'' in \emph{Proceedings of the 20th ACM SIGKDD
  international conference on Knowledge discovery and data mining}, 2014, pp.
  1226--1235.

\bibitem{kinney2005modeling}
R.~Kinney, P.~Crucitti, R.~Albert, and V.~Latora, ``Modeling cascading failures
  in the north american power grid,'' \emph{The European Physical Journal
  B-Condensed Matter and Complex Systems}, vol.~46, no.~1, pp. 101--107, 2005.

\bibitem{klau2005robustness}
G.~W. Klau and R.~Weiskircher, ``Robustness and resilience,'' in \emph{Network
  analysis}.\hskip 1em plus 0.5em minus 0.4em\relax Springer, 2005, pp.
  417--437.

\bibitem{latora2005vulnerability}
V.~Latora and M.~Marchiori, ``Vulnerability and protection of infrastructure
  networks,'' \emph{Physical Review E}, vol.~71, no.~1, p. 015103, 2005.

\bibitem{le2015met}
L.~T. Le, T.~Eliassi-Rad, and H.~Tong, ``Met: A fast algorithm for minimizing
  propagation in large graphs with small eigen-gaps,'' in \emph{Proceedings of
  the 2015 SIAM International Conference on Data Mining}.\hskip 1em plus 0.5em
  minus 0.4em\relax SIAM, 2015, pp. 694--702.

\bibitem{leskovec2007cost}
J.~Leskovec, A.~Krause, C.~Guestrin, C.~Faloutsos, J.~VanBriesen, and
  N.~Glance, ``Cost-effective outbreak detection in networks,'' in
  \emph{Proceedings of the 13th ACM SIGKDD international conference on
  Knowledge discovery and data mining}, 2007, pp. 420--429.

\bibitem{liu2017comparative}
J.~Liu, M.~Zhou, S.~Wang, and P.~Liu, ``A comparative study of network
  robustness measures,'' \emph{Frontiers of Computer Science}, vol.~11, no.~4,
  pp. 568--584, 2017.

\bibitem{lu2016attack}
Z.-M. Lu and X.-F. Li, ``Attack vulnerability of network controllability,''
  \emph{PloS one}, vol.~11, no.~9, p. e0162289, 2016.

\bibitem{malliaros2012fast}
F.~D. Malliaros, V.~Megalooikonomou, and C.~Faloutsos, ``Fast robustness
  estimation in large social graphs: Communities and anomaly detection,'' in
  \emph{Proceedings of the 2012 SIAM International Conference on Data
  Mining}.\hskip 1em plus 0.5em minus 0.4em\relax SIAM, 2012, pp. 942--953.

\bibitem{marzo2018selecting}
J.~L. Marzo, E.~Calle, S.~G. Cosgaya, D.~Rueda, and A.~Ma{\~n}osa, ``On
  selecting the relevant metrics of network robustness,'' in \emph{2018 10th
  International Workshop on Resilient Networks Design and Modeling
  (RNDM)}.\hskip 1em plus 0.5em minus 0.4em\relax IEEE, 2018, pp. 1--7.

\bibitem{mattsson2015vulnerability}
L.-G. Mattsson and E.~Jenelius, ``Vulnerability and resilience of transport
  systems--a discussion of recent research,'' \emph{Transportation Research
  Part A: Policy and Practice}, vol.~81, pp. 16--34, 2015.

\bibitem{mieghem2011decreasing}
P.~Van~Mieghem, D.~Stevanovi{\'c}, F.~Kuipers, C.~Li, R.~Van De~Bovenkamp,
  D.~Liu, and H.~Wang, ``Decreasing the spectral radius of a graph by link
  removals,'' \emph{Physical Review E}, vol.~84, no.~1, p. 016101, 2011.

\bibitem{milanese2010approximating}
A.~Milanese, J.~Sun, and T.~Nishikawa, ``Approximating spectral impact of
  structural perturbations in large networks,'' \emph{Physical Review E},
  vol.~81, no.~4, p. 046112, 2010.

\bibitem{motter2002cascade}
A.~E. Motter and Y.-C. Lai, ``Cascade-based attacks on complex networks,''
  \emph{Physical Review E}, vol.~66, no.~6, p. 065102, 2002.

\bibitem{nardo2018applications}
A.~Di~Nardo, C.~Giudicianni, R.~Greco, M.~Herrera, and G.~F. Santonastaso,
  ``Applications of graph spectral techniques to water distribution network
  management,'' \emph{Water}, vol.~10, no.~1, p.~45, 2018.

\bibitem{nguyen2013detecting}
D.~T. Nguyen, Y.~Shen, and M.~T. Thai, ``Detecting critical nodes in
  interdependent power networks for vulnerability assessment,'' \emph{IEEE
  Transactions on Smart Grid}, vol.~4, no.~1, pp. 151--159, 2013.

\bibitem{parandehgheibi2013robustness}
M.~Parandehgheibi and E.~Modiano, ``Robustness of interdependent networks: The
  case of communication networks and the power grid,'' in \emph{2013 IEEE
  Global Communications Conference (GLOBECOM)}.\hskip 1em plus 0.5em minus
  0.4em\relax IEEE, 2013, pp. 2164--2169.

\bibitem{parshani2010interdependent}
R.~Parshani, S.~V. Buldyrev, and S.~Havlin, ``Interdependent networks: Reducing
  the coupling strength leads to a change from a first to second order
  percolation transition,'' \emph{Physical review letters}, vol. 105, no.~4, p.
  048701, 2010.

\bibitem{paul2004optimization}
G.~Paul, T.~Tanizawa, S.~Havlin, and H.~E. Stanley, ``Optimization of
  robustness of complex networks,'' \emph{The European Physical Journal B},
  vol.~38, no.~2, pp. 187--191, 2004.

\bibitem{prakash2010virus}
B.~A. Prakash, H.~Tong, N.~Valler, M.~Faloutsos, and C.~Faloutsos, ``Virus
  propagation on time-varying networks: Theory and immunization algorithms,''
  in \emph{Joint European Conference on Machine Learning and Knowledge
  Discovery in Databases}.\hskip 1em plus 0.5em minus 0.4em\relax Springer,
  2010, pp. 99--114.

\bibitem{prakash2012threshold}
B.~A. Prakash, D.~Chakrabarti, N.~C. Valler, M.~Faloutsos, and C.~Faloutsos,
  ``Threshold conditions for arbitrary cascade models on arbitrary networks,''
  \emph{Knowledge and information systems}, vol.~33, no.~3, pp. 549--575, 2012.

\bibitem{prakash2013fractional}
B.~A. Prakash, L.~Adamic, T.~Iwashyna, H.~Tong, and C.~Faloutsos, ``Fractional
  immunization in networks,'' in \emph{Proceedings of the 2013 SIAM
  International Conference on Data Mining}.\hskip 1em plus 0.5em minus
  0.4em\relax SIAM, 2013, pp. 659--667.

\bibitem{rueda2017robustness}
D.~F. Rueda, E.~Calle, and J.~L. Marzo, ``Robustness comparison of 15 real
  telecommunication networks: Structural and centrality measurements,''
  \emph{Journal of Network and Systems Management}, vol.~25, no.~2, pp.
  269--289, 2017.

\bibitem{saha2015approximation}
S.~Saha, A.~Adiga, B.~A. Prakash, and A.~K.~S. Vullikanti, ``Approximation
  algorithms for reducing the spectral radius to control epidemic spread,'' in
  \emph{Proceedings of the 2015 SIAM International Conference on Data
  Mining}.\hskip 1em plus 0.5em minus 0.4em\relax SIAM, 2015, pp. 568--576.

\bibitem{schneider2011mitigation}
C.~M. Schneider, A.~A. Moreira, J.~S. Andrade, S.~Havlin, and H.~J. Herrmann,
  ``Mitigation of malicious attacks on networks,'' \emph{Proceedings of the
  National Academy of Sciences}, vol. 108, no.~10, pp. 3838--3841, 2011.

\bibitem{schneider2011suppressing}
C.~M. Schneider, T.~Mihaljev, S.~Havlin, and H.~J. Herrmann, ``Suppressing
  epidemics with a limited amount of immunization units,'' \emph{Physical
  Review E}, vol.~84, no.~6, p. 061911, 2011.

\bibitem{shao2011cascade}
J.~Shao, S.~V. Buldyrev, S.~Havlin, and H.~E. Stanley, ``Cascade of failures in
  coupled network systems with multiple support-dependence relations,''
  \emph{Physical Review E}, vol.~83, no.~3, p. 036116, 2011.

\bibitem{shargel2003optimization}
B.~Shargel, H.~Sayama, I.~R. Epstein, and Y.~Bar-Yam, ``Optimization of
  robustness and connectivity in complex networks,'' \emph{Physical review
  letters}, vol.~90, no.~6, p. 068701, 2003.

\bibitem{sydney2008elasticity}
A.~Sydney, C.~Scoglio, P.~Schumm, and R.~Kooij, ``Elasticity: topological
  characterization of robustness in complex networks,'' \emph{arXiv preprint
  arXiv:0811.4040}, 2008.

\bibitem{tanaka2012dynamical}
G.~Tanaka, K.~Morino, and K.~Aihara, ``Dynamical robustness in complex
  networks: the crucial role of low-degree nodes,'' \emph{Nature}, vol.~2,
  no.~1, pp. 1--6, 2012.

\bibitem{tong2012gelling}
H.~Tong, B.~A. Prakash, T.~Eliassi-Rad, M.~Faloutsos, and C.~Faloutsos,
  ``Gelling, and melting, large graphs by edge manipulation,'' in
  \emph{Proceedings of the 21st ACM international conference on Information and
  knowledge management}, 2012, pp. 245--254.

\bibitem{torres2020node}
L.~Torres, K.~S. Chan, H.~Tong, and T.~Eliassi-Rad, ``Node immunization with
  non-backtracking eigenvalues,'' \emph{arXiv preprint arXiv:2002.12309}, 2020.

\bibitem{trajanovski2013robustness}
S.~Trajanovski, J.~Mart{\'\i}n-Hern{\'a}ndez, W.~Winterbach, and
  P.~Van~Mieghem, ``Robustness envelopes of networks,'' \emph{Journal of
  Complex Networks}, vol.~1, no.~1, pp. 44--62, 2013.

\bibitem{vespignani2010fragility}
A.~Vespignani, ``The fragility of interdependency,'' \emph{Nature}, vol. 464,
  no. 7291, pp. 984--985, 2010.

\bibitem{wang2008attack}
J.~Wang, L.~Rong, L.~Zhang, and Z.~Zhang, ``Attack vulnerability of scale-free
  networks due to cascading failures,'' \emph{Physica A: Statistical Mechanics
  and its Applications}, vol. 387, no.~26, pp. 6671--6678, 2008.

\bibitem{wang2014improving}
X.~Wang, E.~Pournaras, R.~E. Kooij, and P.~Van~Mieghem, ``Improving robustness
  of complex networks via the effective graph resistance,'' \emph{The European
  Physical Journal B}, vol.~87, no.~9, pp. 1--12, 2014.

\bibitem{watts1998collective}
D.~J. Watts and S.~H. Strogatz, ``Collective dynamics of `small-world'
  networks,'' \emph{nature}, vol. 393, no. 6684, pp. 440--442, 1998.

\bibitem{jun2010natural}
W.~Jun, M.~Barahona, T.~Yue-Jin, and D.~Hong-Zhong, ``Natural connectivity of
  complex networks,'' \emph{Chinese physics letters}, vol.~27, no.~7, p.
  078902, 2010.

\bibitem{wu2011spectral}
J.~Wu, M.~Barahona, Y.-J. Tan, and H.-Z. Deng, ``Spectral measure of structural
  robustness in complex networks,'' \emph{IEEE Transactions on Systems, Man,
  and Cybernetics-Part A: Systems and Humans}, vol.~41, no.~6, pp. 1244--1252,
  2011.

\bibitem{xia2010cascading}
Y.~Xia, J.~Fan, and D.~Hill, ``Cascading failure in watts--strogatz small-world
  networks,'' \emph{Physica A: Statistical Mechanics and its Applications},
  vol. 389, no.~6, pp. 1281--1285, 2010.

\bibitem{yang2015improving}
Y.~Yang, Z.~Li, Y.~Chen, X.~Zhang, and S.~Wang, ``Improving the robustness of
  complex networks with preserving community structure,'' \emph{PloS one},
  vol.~10, no.~2, p. e0116551, 2015.

\bibitem{yazdani2010robustness}
A.~Yazdani and P.~Jeffrey, ``Robustness and vulnerability analysis of water
  distribution networks using graph theoretic and complex network principles,''
  in \emph{Water Distribution Systems Analysis 2010}, 2010, pp. 933--945.

\bibitem{yazdani2011complex}
------, ``Complex network analysis of water distribution systems,''
  \emph{Chaos: An Interdisciplinary Journal of Nonlinear Science}, vol.~21,
  no.~1, 2011.

\bibitem{zeng2012enhancing}
A.~Zeng and W.~Liu, ``Enhancing network robustness against malicious attacks,''
  \emph{Physical Review E}, vol.~85, no.~6, p. 066130, 2012.

\bibitem{zhao2004attack}
L.~Zhao, K.~Park, and Y.-C. Lai, ``Attack vulnerability of scale-free networks
  due to cascading breakdown,'' \emph{Physical review E}, vol.~70, no.~3, p.
  035101, 2004.

\bibitem{zhao2014immunization}
D.~Zhao, L.~Wang, S.~Li, Z.~Wang, L.~Wang, and B.~Gao, ``Immunization of
  epidemics in multiplex networks,'' \emph{PloS one}, vol.~9, no.~11, p.
  e112018, 2014.

\bibitem{jung1978class}
H.~A. Jung, ``On a class of posets and the corresponding comparability
  graphs,'' \emph{Journal of Combinatorial Theory, Series B}, vol.~24, no.~2,
  pp. 125--133, 1978.

\bibitem{cozzens1995tenacity}
M.~Cozzens, D.~Moazzami, and S.~Stueckle, ``The tenacity of a graph,'' 1995.

\bibitem{barefoot1987integrity}
C.~Barefoot, R.~Entringer, and H.~Swart, ``Integrity of trees and powers of
  cycles,'' \emph{Congr. Numer}, vol.~58, pp. 103--114, 1987.

\bibitem{mohar1989isoperimetric}
B.~Mohar, ``Isoperimetric numbers of graphs,'' \emph{Journal of combinatorial
  theory, Series B}, vol.~47, no.~3, pp. 274--291, 1989.

\bibitem{henzinger2000computing}
M.~R. Henzinger, S.~Rao, and H.~N. Gabow, ``Computing vertex connectivity: New
  bounds from old techniques,'' \emph{Journal of Algorithms}, vol.~34, no.~2,
  pp. 222--250, 2000.

\bibitem{matula1987determining}
D.~W. Matula, ``Determining edge connectivity in 0 (nm),'' in \emph{28th Annual
  Symposium on Foundations of Computer Science (sfcs 1987)}.\hskip 1em plus
  0.5em minus 0.4em\relax IEEE, 1987, pp. 249--251.

\bibitem{esfahanian2013connectivity}
A.-H. Esfahanian, ``Connectivity algorithms,'' in \emph{Topics in structural
  graph theory}.\hskip 1em plus 0.5em minus 0.4em\relax Cambridge University
  Press, 2013, pp. 268--281.

\bibitem{whitney1992congruent}
H.~Whitney, ``Congruent graphs and the connectivity of graphs,'' in
  \emph{Hassler Whitney Collected Papers}.\hskip 1em plus 0.5em minus
  0.4em\relax Springer, 1992, pp. 61--79.

\bibitem{warshall1962algorithm}
\BIBentryALTinterwordspacing
R.~W. Floyd, ``Algorithm 97: Shortest path,'' \emph{Commun. ACM}, vol.~5,
  no.~6, p. 345, Jun. 1962. [Online]. Available:
  \url{https://doi.org/10.1145/367766.368168}
\BIBentrySTDinterwordspacing

\bibitem{freeman1977set}
L.~C. Freeman, ``A set of measures of centrality based on betweenness,''
  \emph{Sociometry}, pp. 35--41, 1977.

\bibitem{brandes2001faster}
U.~Brandes, ``A faster algorithm for betweenness centrality,'' \emph{Journal of
  mathematical sociology}, vol.~25, no.~2, pp. 163--177, 2001.

\bibitem{green2013faster}
O.~Green and D.~A. Bader, ``Faster clustering coefficient using vertex
  covers,'' in \emph{2013 International Conference on Social Computing}.\hskip
  1em plus 0.5em minus 0.4em\relax IEEE, 2013, pp. 321--330.

\bibitem{butler2008eigenvalues}
S.~K. Butler, ``Eigenvalues and structures of graphs,'' Ph.D. dissertation, UC
  San Diego, 2008.

\bibitem{wang2003epidemic}
Y.~Wang, D.~Chakrabarti, C.~Wang, and C.~Faloutsos, ``Epidemic spreading in
  real networks: An eigenvalue viewpoint,'' in \emph{22nd International
  Symposium on Reliable Distributed Systems, 2003. Proceedings.}\hskip 1em plus
  0.5em minus 0.4em\relax IEEE, 2003, pp. 25--34.

\bibitem{chen2015fast}
C.~Chen and H.~Tong, ``Fast eigen-functions tracking on dynamic graphs,'' in
  \emph{Proceedings of the 2015 SIAM International Conference on Data
  Mining}.\hskip 1em plus 0.5em minus 0.4em\relax SIAM, 2015, pp. 559--567.

\bibitem{estrada2005subgraph}
E.~Estrada and J.~A. Rodriguez-Velazquez, ``Subgraph centrality in complex
  networks,'' \emph{Physical Review E}, vol.~71, no.~5, p. 056103, 2005.

\bibitem{hoory2006expander}
S.~Hoory, N.~Linial, and A.~Wigderson, ``Expander graphs and their
  applications,'' \emph{Bulletin of the American Mathematical Society},
  vol.~43, no.~4, pp. 439--561, 2006.

\bibitem{estrada2005spectral}
E.~Estrada and J.~A. Rodr{\'\i}guez-Vel{\'a}zquez, ``Spectral measures of
  bipartivity in complex networks,'' \emph{Physical Review E}, vol.~72, no.~4,
  p. 046105, 2005.

\bibitem{wu2018scalable}
L.~Wu, P.-Y. Chen, I.~E.-H. Yen, F.~Xu, Y.~Xia, and C.~Aggarwal, ``Scalable
  spectral clustering using random binning features,'' in \emph{Proceedings of
  the 24th ACM SIGKDD International Conference on Knowledge Discovery \& Data
  Mining}, 2018, pp. 2506--2515.

\bibitem{fiedler1973algebraic}
M.~Fiedler, ``Algebraic connectivity of graphs,'' \emph{Czechoslovak
  mathematical journal}, vol.~23, no.~2, pp. 298--305, 1973.

\bibitem{lehoucq1998arpack}
R.~B. Lehoucq, D.~C. Sorensen, and C.~Yang, \emph{ARPACK users' guide: solution
  of large-scale eigenvalue problems with implicitly restarted Arnoldi
  methods}.\hskip 1em plus 0.5em minus 0.4em\relax Siam, 1998, vol.~6.

\bibitem{weichenberg2004high}
G.~Weichenberg, V.~W. Chan, and M.~M{\'e}dard, ``High-reliability topological
  architectures for networks under stress,'' \emph{IEEE Journal on Selected
  Areas in Communications}, vol.~22, no.~9, pp. 1830--1845, 2004.

\bibitem{buekenhout1998number}
F.~Buekenhout and M.~Parker, ``The number of nets of the regular convex
  polytopes in dimension < 4,'' \emph{Discrete mathematics}, vol. 186, no. 1-3,
  pp. 69--94, 1998.

\bibitem{tsironis2013accurate}
S.~Tsironis, M.~Sozio, M.~Vazirgiannis, and L.~Poltechnique, ``Accurate
  spectral clustering for community detection in mapreduce,'' in \emph{Advances
  in Neural Information Processing Systems (NIPS) Workshops}.\hskip 1em plus
  0.5em minus 0.4em\relax Citeseer, 2013.

\bibitem{lordan2016robustness}
O.~Lordan, J.~M. Sallan, N.~Escorihuela, and D.~Gonzalez-Prieto, ``Robustness
  of airline route networks,'' \emph{Physica A: Statistical Mechanics and its
  Applications}, vol. 445, pp. 18--26, 2016.

\bibitem{du2016analysis}
W.-B. Du, X.-L. Zhou, O.~Lordan, Z.~Wang, C.~Zhao, and Y.-B. Zhu, ``Analysis of
  the chinese airline network as multi-layer networks,'' \emph{Transportation
  Research Part E: Logistics and Transportation Review}, vol.~89, pp. 108--116,
  2016.

\bibitem{lordan2015robustness}
O.~Lordan, J.~M. Sallan, P.~Simo, and D.~Gonzalez-Prieto, ``Robustness of
  airline alliance route networks,'' \emph{Communications in Nonlinear Science
  and Numerical Simulation}, vol.~22, no. 1-3, pp. 587--595, 2015.

\bibitem{zhou2019efficiency}
Y.~Zhou, J.~Wang, and G.~Q. Huang, ``Efficiency and robustness of weighted air
  transport networks,'' \emph{Transportation Research Part E: Logistics and
  Transportation Review}, vol. 122, pp. 14--26, 2019.

\bibitem{berche2009resilience}
B.~Berche, C.~Von~Ferber, T.~Holovatch, and Y.~Holovatch, ``Resilience of
  public transport networks against attacks,'' \emph{The European Physical
  Journal B}, vol.~71, no.~1, pp. 125--137, 2009.

\bibitem{nguyen2021modularity}
Q.~Nguyen, T.~V. Vu, H.-D. Dinh, D.~Cassi, F.~Scotognella, R.~Alfieri, and
  M.~Bellingeri, ``Modularity affects the robustness of scale-free model and
  real-world social networks under betweenness and degree-based node attack,''
  \emph{Applied Network Science}, vol.~6, no.~1, pp. 1--21, 2021.

\bibitem{boldi2011robustness}
P.~Boldi, M.~Rosa, and S.~Vigna, ``Robustness of social networks: Comparative
  results based on distance distributions,'' in \emph{International Conference
  on Social Informatics}.\hskip 1em plus 0.5em minus 0.4em\relax Springer,
  2011, pp. 8--21.

\bibitem{moore2021inclusivity}
J.~M. Moore, M.~Small, and G.~Yan, ``Inclusivity enhances robustness and
  efficiency of social networks,'' \emph{Physica A: Statistical Mechanics and
  its Applications}, vol. 563, p. 125490, 2021.

\bibitem{colladon2017robustness}
A.~F. Colladon and F.~Vagaggini, ``Robustness and stability of enterprise
  intranet social networks: The impact of moderators,'' \emph{Information
  Processing \& Management}, vol.~53, no.~6, pp. 1287--1298, 2017.

\bibitem{boldi2013robustness}
P.~Boldi, M.~Rosa, and S.~Vigna, ``Robustness of social and web graphs to node
  removal,'' \emph{Social Network Analysis and Mining}, vol.~3, no.~4, pp.
  829--842, 2013.

\bibitem{bilal2013characterization}
K.~Bilal, M.~Manzano, S.~U. Khan, E.~Calle, K.~Li, and A.~Y. Zomaya, ``On the
  characterization of the structural robustness of data center networks,''
  \emph{IEEE Transactions on Cloud Computing}, vol.~1, no.~1, pp. 1--1, 2013.

\bibitem{manzano2013connectivity}
M.~Manzano, K.~Bilal, E.~Calle, and S.~U. Khan, ``On the connectivity of data
  center networks,'' \emph{IEEE Communications Letters}, vol.~17, no.~11, pp.
  2172--2175, 2013.

\bibitem{de2016reliability}
R.~de~Souza~Couto, S.~Secci, M.~E.~M. Campista, and L.~H. M.~K. Costa,
  ``Reliability and survivability analysis of data center network topologies,''
  \emph{Journal of Network and Systems Management}, vol.~24, no.~2, pp.
  346--392, 2016.

\bibitem{cohen2011resilience}
R.~Cohen, K.~Erez, S.~Havlinl, M.~Newman, A.-L. Barab{\'a}si, D.~J. Watts
  \emph{et~al.}, ``Resilience of the internet to random breakdowns,'' in
  \emph{The Structure and Dynamics of Networks}.\hskip 1em plus 0.5em minus
  0.4em\relax Princeton University Press, 2011, pp. 507--509.

\bibitem{di2018applications}
A.~Di~Nardo, C.~Giudicianni, R.~Greco, M.~Herrera, and G.~F. Santonastaso,
  ``Applications of graph spectral techniques to water distribution network
  management,'' \emph{Water}, vol.~10, no.~1, p.~45, 2018.

\bibitem{ebel2002scale}
H.~Ebel, L.-I. Mielsch, and S.~Bornholdt, ``Scale-free topology of e-mail
  networks,'' \emph{Physical review E}, vol.~66, no.~3, p. 035103, 2002.

\bibitem{faloutsos2011power}
M.~Faloutsos, P.~Faloutsos, and C.~Faloutsos, ``On power-law relationships of
  the internet topology,'' in \emph{The Structure and Dynamics of
  Networks}.\hskip 1em plus 0.5em minus 0.4em\relax Princeton University Press,
  2011, pp. 195--206.

\bibitem{jeong2000large}
H.~Jeong, B.~Tombor, R.~Albert, Z.~N. Oltvai, and A.-L. Barab{\'a}si, ``The
  large-scale organization of metabolic networks,'' \emph{Nature}, vol. 407,
  no. 6804, pp. 651--654, 2000.

\bibitem{mislove2007measurement}
A.~Mislove, M.~Marcon, K.~P. Gummadi, P.~Druschel, and B.~Bhattacharjee,
  ``Measurement and analysis of online social networks,'' in \emph{Proceedings
  of the 7th ACM SIGCOMM conference on Internet measurement}, 2007, pp. 29--42.

\bibitem{albert2005scale}
R.~Albert, ``Scale-free networks in cell biology,'' \emph{Journal of cell
  science}, vol. 118, no.~21, pp. 4947--4957, 2005.

\bibitem{clauset2009power}
A.~Clauset, C.~R. Shalizi, and M.~E. Newman, ``Power-law distributions in
  empirical data,'' \emph{SIAM review}, vol.~51, no.~4, pp. 661--703, 2009.

\bibitem{akoglu2009rtg}
L.~Akoglu and C.~Faloutsos, ``Rtg: A recursive realistic graph generator using
  random typing,'' in \emph{Joint European Conference on Machine Learning and
  Knowledge Discovery in Databases}.\hskip 1em plus 0.5em minus 0.4em\relax
  Springer, 2009, pp. 13--28.

\bibitem{erdos1960evolution}
P.~Erd{\H{o}}s and A.~R{\'e}nyi, ``On the evolution of random graphs,''
  \emph{Publ. Math. Inst. Hung. Acad. Sci}, vol.~5, no.~1, pp. 17--60, 1960.

\bibitem{chakrabarti2006graph}
D.~Chakrabarti and C.~Faloutsos, ``Graph mining: Laws, generators, and
  algorithms,'' \emph{ACM computing surveys (CSUR)}, vol.~38, no.~1, pp. 2--es,
  2006.

\bibitem{barabasi1999emergence}
A.-L. Barab{\'a}si and R.~Albert, ``Emergence of scaling in random networks,''
  \emph{science}, vol. 286, no. 5439, pp. 509--512, 1999.

\bibitem{holme2002growing}
P.~Holme and B.~J. Kim, ``Growing scale-free networks with tunable
  clustering,'' \emph{Physical review E}, vol.~65, no.~2, p. 026107, 2002.

\bibitem{page1999pagerank}
L.~Page, S.~Brin, R.~Motwani, and T.~Winograd, ``The pagerank citation ranking:
  Bringing order to the web.'' Stanford InfoLab, Tech. Rep., 1999.

\bibitem{opsahl2010node}
T.~Opsahl, F.~Agneessens, and J.~Skvoretz, ``Node centrality in weighted
  networks: Generalizing degree and shortest paths,'' \emph{Social networks},
  vol.~32, no.~3, pp. 245--251, 2010.

\bibitem{katz1953new}
L.~Katz, ``A new status index derived from sociometric analysis,''
  \emph{Psychometrika}, vol.~18, no.~1, pp. 39--43, 1953.

\bibitem{kang2018aurora}
J.~Kang, M.~Wang, N.~Cao, Y.~Xia, W.~Fan, and H.~Tong, ``Aurora: Auditing
  pagerank on large graphs,'' in \emph{2018 IEEE International Conference on
  Big Data (Big Data)}.\hskip 1em plus 0.5em minus 0.4em\relax IEEE, 2018, pp.
  713--722.

\bibitem{wang2020graph}
M.~Wang, J.~Kang, C.~Nan, Y.~Xia, W.~Fan, and H.~Tong, ``Graph ranking
  auditing: Problem definition and fast solutions,'' \emph{IEEE Transactions on
  Knowledge and Data Engineering}, 2020.

\bibitem{kang2018x}
J.~Kang, S.~Freitas, H.~Yu, Y.~Xia, N.~Cao, and H.~Tong, ``X-rank: Explainable
  ranking in complex multi-layered networks,'' in \emph{Proceedings of the 27th
  ACM International Conference on Information and Knowledge Management}, 2018,
  pp. 1959--1962.

\bibitem{xie2020auditing}
T.~Xie, Y.~Ma, H.~Tong, M.~T. Thai, and R.~Maciejewski, ``Auditing the
  sensitivity of graph-based ranking with visual analytics,'' \emph{IEEE
  Transactions on Visualization and Computer Graphics}, 2020.

\bibitem{chen2018realtime}
Z.~Chen, H.~Tong, and L.~Ying, ``Realtime robustification of interdependent
  networks under cascading attacks,'' in \emph{2018 IEEE International
  Conference on Big Data (Big Data)}.\hskip 1em plus 0.5em minus 0.4em\relax
  IEEE, 2018, pp. 1347--1356.

\bibitem{danziger2016effect}
M.~M. Danziger, L.~M. Shekhtman, Y.~Berezin, and S.~Havlin, ``The effect of
  spatiality on multiplex networks,'' \emph{EPL (Europhysics Letters)}, vol.
  115, no.~3, p. 36002, 2016.

\bibitem{barthelemy2011spatial}
M.~Barth{\'e}lemy, ``Spatial networks,'' \emph{Physics Reports}, vol. 499, no.
  1-3, pp. 1--101, 2011.

\end{thebibliography}

% \begin{IEEEbiography}[{\includegraphics[width=1in,height=1.25in,clip,keepaspectratio]{mshell}}]{Michael Shell}

\vfill\break
\begin{IEEEbiography}[{\includegraphics[width=1in,height=1.25in,clip,keepaspectratio]{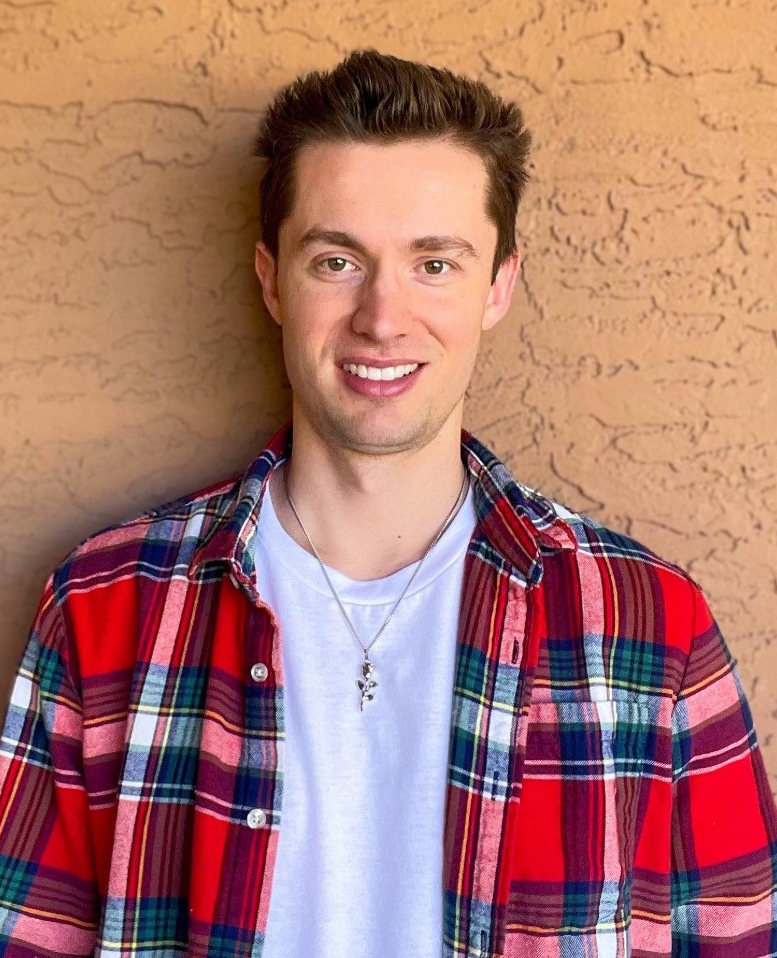}}]{Scott Freitas} is an applied scientist at Microsoft working to develop explainable, robust, and efficient next-generation cybersecurity machine learning systems.
He received his machine learning PhD from Georgia Institute of Technology in the department of Computational Science and Engineering.
His work has been supported by fellowships from IBM Research, NSF GRFP, and Raytheon; and he has co-authored several winning research proposals, including a multi-million dollar DARPA grant.

\end{IEEEbiography}

\begin{IEEEbiography}[{\includegraphics[width=1in,height=1.25in,clip,keepaspectratio]{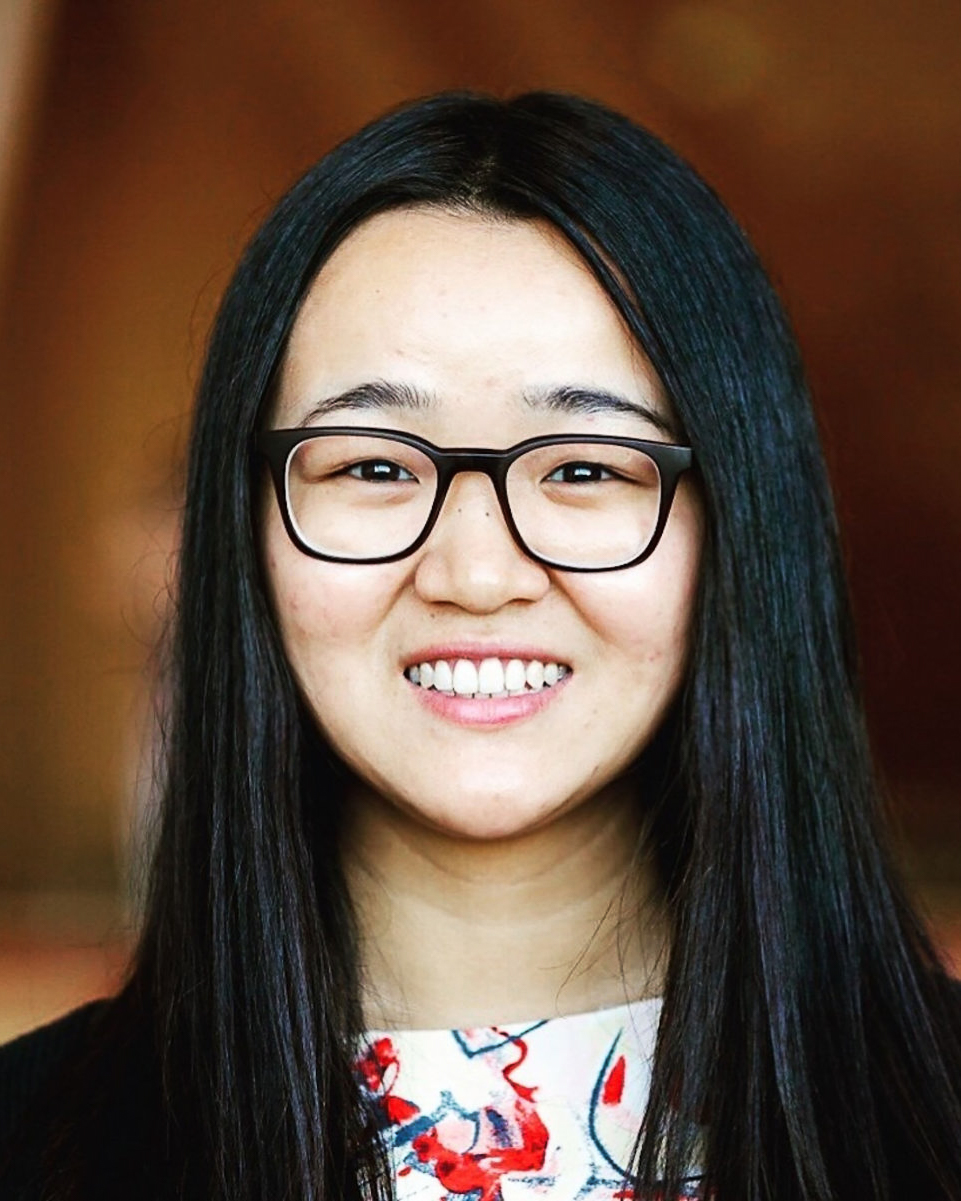}}]{Diyi Yang}
is an Assistant Professor at Georgia Tech.
She is broadly interested in computational social science, natural language processing, and machine learning.  
Diyi received her PhD from the Language Technologies Institute at Carnegie Mellon University. Her work has been published at leading NLP/HCI conferences, and also resulted in multiple best paper award nominations. Diyi is named among IEEE AI's 10 to Watch in 2020, and Forbes 30 under 30 in Science.
\end{IEEEbiography}

%\newpage

\begin{IEEEbiography}[{\includegraphics[width=1in,height=1.25in,clip,keepaspectratio]{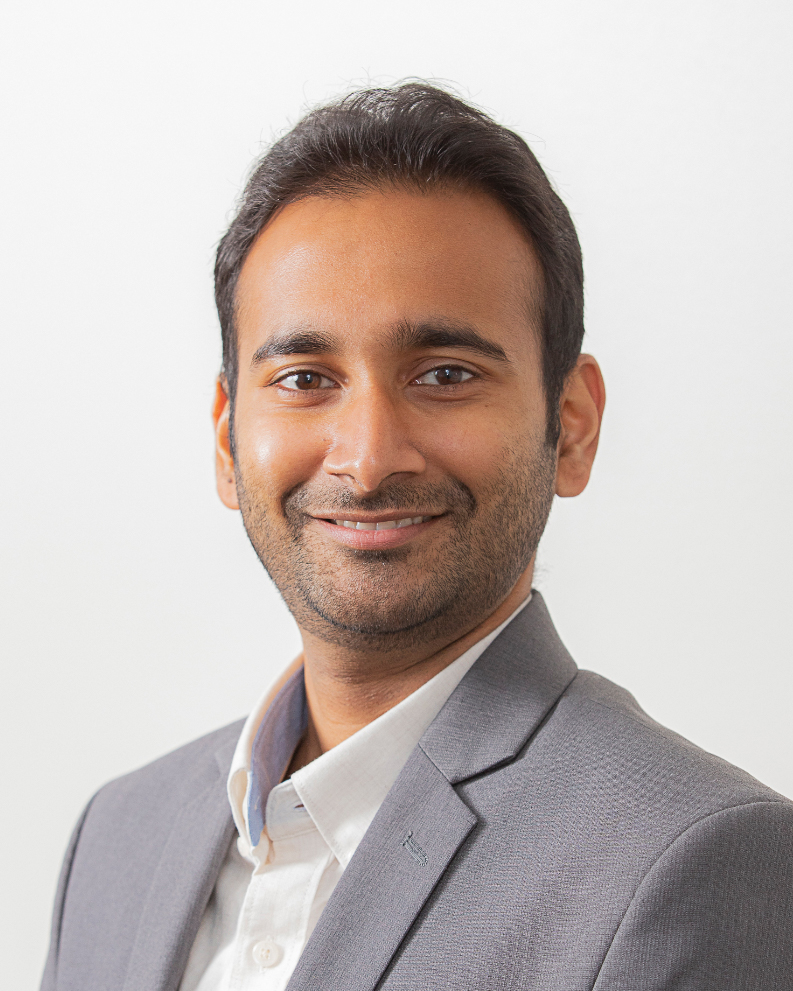}}]{Srijan Kumar} is an Assistant Professor at College of Computing at Georgia Institute of Technology.
He is broadly interested in data science and applied machine learning solutions to improve safety, integrity, and well-being in the web and society.
He is the recipient of the Facebook Faculty Research Award, Adobe Faculty Research Award, SIGKDD Doctoral Dissertation Award 2018 runner-up, WWW 2017 Best Paper Award runner-up, Best of ICDM 2016, and Larry S. Davis Doctoral Dissertation Award 2017.
\end{IEEEbiography}

\begin{IEEEbiography}[{\includegraphics[width=1in,height=1.25in,clip,keepaspectratio]{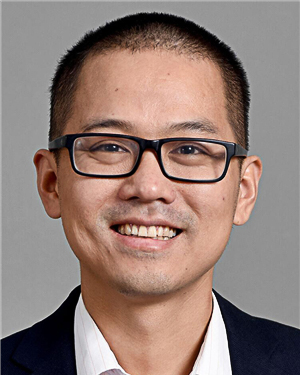}}]{Hanghang Tong} 
is an associate professor in the Department of Computer Science at University of Illinois at Urbana-Champaign. 
Before that, he was an assistant professor at
Arizona State University; an assistant professor in 
City University of New York; and a research staff member at IBM T.J. Watson Research Center.
His research interest includes large scale data mining for graphs and multimedia. He has received several awards, including Best Paper Award in CIKM 2012, SDM 2008, and ICDM 2006. 
\end{IEEEbiography}

\begin{IEEEbiography}[{\includegraphics[width=1in,height=1.25in,clip,keepaspectratio]{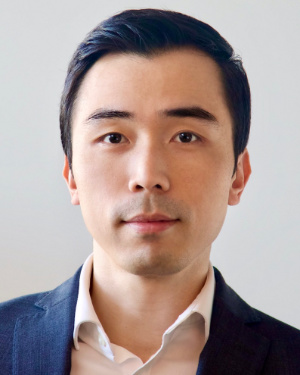}}]{Duen Horng (Polo) Chau} 
is an Associate Professor at Georgia Tech. 
His research bridges data mining and HCI to make sense of massive
datasets. 
His thesis won Carnegie Mellon’s CS Dissertation Award, Honorable Mention. 
He received awards from Intel, Google, Yahoo, LexisNexis, and Symantec; he won paper awards
at SIGMOD, KDD and SDM. 
He is an ACM IUI steering committee member, IUI15 co-chair, and
IUI19 program co-chair. 
His research is deployed by Facebook, Symantec, and Yahoo.
\end{IEEEbiography}

% \vfill


\end{document}